\documentclass[draftcls,onecolumn,twoside,letter,12 pt ]{IEEEtran}
\usepackage{graphicx,subfigure}
\usepackage{caption}
\usepackage[cmex10]{amsmath}
\newcounter{subeqn} %
\makeatletter
\@addtoreset{subeqn}{equation}
\makeatother

\usepackage{algorithmic}
\usepackage{algorithm}
\usepackage{amssymb}
\usepackage{mathrsfs}
\usepackage{stfloats}
\usepackage{dsfont}
\usepackage{setspace}
\usepackage{epstopdf}
\usepackage{epsfig}
\usepackage{balance}
\usepackage{enumitem}
\usepackage{pstricks}
\usepackage{hyperref}
\usepackage[noadjust]{cite}
\usepackage{filecontents}
\usepackage{multirow}

\renewcommand{\citedash}{--}

\newtheorem{definition}{Definition}
\newtheorem{theorem}{Theorem}

\newtheorem{proposition}{Proposition}

\begin{document}
	\title{Edge-Caching Wireless Networks: Performance Analysis and Optimization}


	\author{\IEEEauthorblockN{Thang~X.~Vu, \IEEEmembership{Member, IEEE}, Symeon~Chatzinotas,  \IEEEmembership{Senior Member, IEEE}, and Bjorn~Ottersten, \IEEEmembership{Fellow, IEEE}}
	\thanks{This research is supported, in part, by the ERC AGNOSTIC project under code R-AGR-3283 and the FNR CORE ProCAST project. }
	\thanks{The authors are with the Interdisciplinary Centre for Security, Reliability and Trust (SnT) -- University of Luxembourg, 29 Avenue John F. Kennedy,
		L-1855 Luxembourg. E--Mail: \{thang.vu,symeon.chatzinotas, bjorn.ottersten\}@uni.lu.}
	\thanks{Parts of this work have been presented at the IEEE SPAWC 2017 \cite{Vu2017a}.}
	}

	\providecommand{\keywords}[1]{\textbf{\textit{Index terms---}} #1}
	
	\date{}
	
	\maketitle
	\thispagestyle{plain}
	\begin{abstract}
		Edge-caching has received much attention as an efficient technique to reduce delivery latency and network congestion during peak-traffic times by bringing data closer to end users. Existing works usually design caching algorithms separately from physical layer design. In this paper, we analyse edge-caching wireless networks by taking into account the caching capability when designing the signal transmission. Particularly, we investigate multi-layer caching where both base station (BS) and users are capable of storing content data in their local cache and analyse the performance of edge-caching wireless networks under two notable uncoded and coded caching strategies. Firstly, we calculate backhaul and access throughputs of the two caching strategies for arbitrary values of cache size. The required backhaul and access throughputs are derived as a function of the BS and user cache sizes. Secondly, closed-form expressions for the system energy efficiency (EE) corresponding to the two caching methods are derived. Based on the derived formulas, the system EE is maximized via precoding vectors design and optimization while satisfying a predefined user request rate. Thirdly, two optimization problems are proposed to minimize the content delivery time for the two caching strategies. Finally, numerical results are presented to verify the effectiveness of the two caching methods. 
	\end{abstract}
	
	\section{Introduction}
Future wireless networks will have to address stringent requirements of delivering content at high speed and low latency due to the proliferation of mobile devices and data-hungry applications. It is predicted that by 2020, more than 70\% of network traffic will be video \cite{Cisco}. 
Although various network architectures have been proposed in order to boost the network throughput and reduce transmission latency such as cloud radio access networks (C-RANs) \cite{Vu2016,Tran1,Tran2} and heterogeneous networks (HetNets), traffic congestion might occur during peak-traffic times. A promising solution to reduce latency and network costs of content delivery is to bring the content closer to end users via distributed storages through out the network, which is referred to content placement or caching \cite{Borst2010}. Caching usually consists of a placement phase and a delivery phase. The former is executed during off-peak periods when the network resources are abundant. In this phase, popular content is stored in the distributed caches. The later usually occurs during peak-traffic times when the actual users' requests are revealed. If the requested content is available in the user's local storage, it can be served immediately without being sent via the network. In this manner, caching allows significant backhaul's load reduction during peak-traffic times and thus mitigating network congestion \cite{Borst2010,AliNie2014}. 
	
Most research works on caching exploit historic user requested data to optimize either placement or delivery phases \cite{Almeroth1996,Borst2010,Christop2015d}. For a fixed content delivery strategy, the placement phase is designed to maximize the local caching gain, which is proportional to the number of file parts available in the local storage. This caching method stores the contents independently and are known as \emph{uncoded} caching. The caching gain can be further improved via multicasting a combination of the requested files during the delivery phase, which is known as \emph{coded} caching \cite{AliNie2014,Vu2017}. By carefully placing the files in the caches and designing the coded data, all users can recover their desired content via a multicast stream. Rate-memory tradeoff is derived in \cite{AliNie2014}, which achieves a global caching gain on top of the local caching gain. This gain is inversely proportional to the total cache memory. Similar rate-memory tradeoff is investigated in device-to-device (D2D) networks \cite{JiCaire2016} and secrecy constraints \cite{Sengupta2016}. In \cite{Sengupta2016b,SHPark2017}, the authors study the tradeoff between the memory at edge nodes and the transmission latency measured in normalized delivery time. The rate-memory tradeoff of multi-layer coded caching networks is studied in \cite{Karamchandani2016,Tang2016}. Note that the global gain brought by the coded caching comes at a price of coordination since the data centre needs to know the number of users in order to construct the coded messages. 

Recently, there have been numerous works addressing joint content caching and transmission design for cache-assisted wireless networks. The main idea is to take into account the cached content at the edge nodes when designing the link transmission to reduce the access and backhaul costs. It is shown in \cite{TaoCheZhoYu2016} that transmit power and fronthaul bandwidth can be reduced via cache-aware multicast beamforming design and power allocation. The impact of wireless backhaul on the energy consumption was studied in \cite{Vu2018a}. The authors in \cite{Khreishah2016} propose a joint optimization of caching, routing and channel assignment via two sub-problems called restricted master and pricing. The performance of caching wireless D2D networks are analysed in \cite{ZhaXiaWuLi2016,Gregori2016,Ji2016,Ji2015a}. In \cite{Gregori2016}, the authors study D2D networks which allow the storage of files at either small base stations or user terminals. Taking into account the wireless fading channels, a joint content replacement and delivering scheme is proposed to reduce the system energy consumption. The throughput-outage trade-off of the mmWave underlying D2D networks under a simplified grid topology is derived in \cite{Ji2016}. The stochastic performance of caching wireless networks is analysed in \cite{Yang2016}, in which the nodes' locations are modelled as a Poison point process (PPP). The average ergodic downlink rate and outage probability are studied when cache capability is present at three tiers of base station (BS), relay and D2D pairs. In \cite{Chen2016}, success delivery rate is studied in  cluster-centric networks, which group small base stations (SBSs) into disjoint clusters. In this work, the SBSs within one cluster share a cache which is divided into two parts: one contains the most popular files, and one comprises different files which are most popular locally. The authors in \cite{Alfano2016} study effects of mobility on the caching wireless networks via a random-walk assumption of node mobilities. In \cite{Tran2016}, a low-complexity greedy algorithm is proposed to minimize the content delivering delay in cooperative caching C-RANs. Energy efficiency (EE) of cache-assisted networks are analysed in \cite{Gabry2016,Liu2016}. Focusing on the content placement phase in heterogeneous networks, the authors in \cite{Gabry2016} study the trade-off between the expected backhaul rate and energy consumption. The impact of caching is analysed in \cite{Liu2016} via close-form expression of the approximated network EE. We note that these works consider either only the uncoded caching method or the caching at higher layers separated from the signal transmission. 

In this paper, we investigate the performance of edge-caching wireless networks in which multi-layer caches are available at either user or edge nodes. Our contributions are as follows:
\begin{itemize}
	\item Firstly, we investigate the performance of edge-caching networks under two notable \emph{uncoded} and inter-file \emph{coded} caching strategies\footnote{The inter-file coded caching is different from intra-file coded caching method.}. In particular, we compute the required throughputs on the backhaul and access links for both caching strategies with arbitrary cache sizes. 
	
	\item Secondly, we derive a closed-form expression for the system EE, which reveals insight contributions of cache capability at the BS and users. Based on the derived formula, we maximize the system EE subject to a quality-of-service (QoS) constraint taking into account the caching strategies. The maximum EE is obtained in closed-form for zero-forcing (ZF) precoding and suboptimally solved via semi-definite relaxation (SDR) design. Our paper differs from \cite{Liu2016,Gabry2016} as following. We focus on the delivery phase, while \cite{Liu2016} considers the placement phase. We consider multi-layer cache and the two caching strategies, while \cite{Gabry2016} only considers caching available at the BS with an uncoded caching algorithm.
	\item Thirdly, we analyse and minimize the delivery time for the two caching strategies via two formulated problems which jointly optimize the beamforming design and power allocation. Our method is fundamentally different from \cite{Sengupta2016b} which studies the latency limit from information-theoretic perspectives. Compared with \cite{Tran2016}, which studies only uncoded caching at higher layers, we consider both caching strategies jointly with the signal transmission. 
	\item Finally, the analysed EE and delivery time are verified via selective numerical results. We show an interesting result that the uncoded-caching is more energy-efficient only for the small user cache sizes. This result is different from the common understanding that the coded caching always outperforms the uncoded caching in terms of total backhaul's throughput.
\end{itemize}
	
The rest of this paper is organised as follows. Section~\ref{sec:SystemModel} presents the system model and the caching strategies. Section~\ref{sec:EE} analyses the system energy efficiency. Section~\ref{sec:EE maximization} presents the proposed EE maximization algorithms. Section~\ref{sec:delivery time} minimizes the delivery time. Section~\ref{sec:Zipf} derives the EE for general content popularity. Section~\ref{sec:Results} shows numerical results. Finally, Section~\ref{sec:Conclusions} concludes the paper.

\emph{Notation}: $(.)^H, (x)^+$ and $\mathrm{Tr}(.)$ denote the Hermitian transpose, $\max(0, x)$ and the $\mathrm{trace}(.)$ function, respectively. $\lfloor x \rfloor$ denotes the largest integer not exceeding $x$.

\section{System Model} \label{sec:SystemModel}
\begin{figure}
	\centering
	\includegraphics[width = \columnwidth]{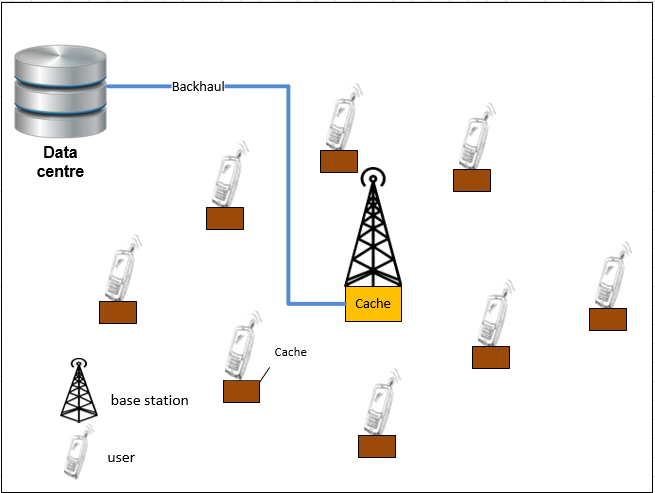}
	\caption{Multiple-layer cache-assisted wireless networks.} \label{fig:SimpleModel}
\end{figure}
We consider the downlink edge-caching wireless network in which a data centre serves $K$ distributed users, denoted by $\mathcal{K} = \{1, \dots, K\}$, via one BS, as depicted in Figure~\ref{fig:SimpleModel}. This model can also be applied in various practical scenarios in which the users can be replaced by various cache-assisted edge nodes, e.g., edge nodes in fog radio access networks (F-RAN), small-cell BSs in HetNet. The $L$-antenna BS, with $L \geq K$, serves all users via wireless access networks and connects to the data centre via an error-free, bandwidth-limited backhaul link. The wireless transmissions are subjected to block Rayleigh fading channels, in which the channel fading coefficients are fixed within a block and are mutually independent across the users. The block duration is assumed to be long enough for the users to be served the requested files. The data centre contains $N$ files of equal size of $Q$ bits and is denoted by $\mathcal{F} = \{F_1, \dots, F_N\}$. In practice, unequal size files can be divided into trunks of subfiles which have the same size. 
	
\subsection{Caching model}
We consider multiple-layer caching networks in which both the BS and users are equipped with a storage memory of size $M_b$ and $M_u$ files, with $0\leq M_b, M_u \leq N$, respectively. We consider off-line caching, in which the \emph{content placement phase} is executed during off-peak times \cite{AliNie2014}. For robustness, we consider the completely distributed placement phase in which the BS is unaware of user cache's content. In particular, the BS stores $\frac{M_bQ}{N}$ (non-overlapping) bits of every file in its cache, which are randomly chosen\footnote{There may exists a better cache placement at the BS at the expense of coordination.}. Similarly, each user stores $\frac{M_uQ}{N}$ bits of every file in its cache under the uncoded caching strategy\footnote{If $\frac{M_bQ}{N}$ or $\frac{M_uQ}{N}$ is not an integer, we round up this ratio to the closest integer and perform zero-padding to the last.}. The placement phase at the user caches under the coded caching is similar to \cite{AliNie2014}.  The total number of bits stored at the BS and user caches are respectively $M_bQ$ and $M_uQ$ bits, which satisfy the memory constraints. 
	
At the beginning of the \emph{delivery phase}, each user requests one file from the library. In order to focus on the interplay between the EE and cache capabilities, we consider the worst case in which the users tend to request different files and the content popularity follows a uniform distribution \cite{AliNie2014}. The general case of content popularities, e.g., Zipf distribution, will be studied in Section~\ref{sec:Zipf}. Denote $d_1, ..., d_K$ as the file indices requested by user $1, ..., K$, respectively. If the requested bits (or subfile) is in its own cache, they can be served immediately. Otherwise, this subfile is sent from the BS's cache or the data centre through the backhaul link.	
We consider two notable caching methods for the delivery phase: uncoded caching and coded caching. 
	\subsubsection{Uncoded caching}
	This strategy sends parts of the requested files to each user independently. We note that the users do not know the cache content of each other. The advantage of this method is the robustness and it does not require coordination. The total number of  bits transmitted through the backhaul link, $Q_{\rm unc, BH}$, and the access link, $Q_{\rm unc, AC}$, are given in the following proposition.
	%
	\begin{proposition}\label{prop:1}
		Under the uncoded caching strategy, the total number of bits transmitted through the backhaul and access links are 
$
		Q_{\rm unc, BH} = KQ\left(1 - \frac{M_u}{N}\right)\left(1 -\frac{M_b}{N} \right)$
		and 
$
		Q_{\rm unc, AC} = KQ\left(1 - \frac{M_u}{N}\right)$,
respectively.
	\end{proposition}
	\begin{IEEEproof}
	See Appendix~\ref{app:1}.
	\end{IEEEproof}
		
	\subsubsection{Coded caching} In coded caching strategy, the data centre first intelligently encodes the requested files and then sends them to the users. We note that this strategy requires the number of users in order to construct the coded messages for all users. 

	\begin{proposition}\label{prop:2}
		Let $m = \lfloor \frac{KM_u}{N} \rfloor \in \mathbb{Z}^\star$ and $\delta = \frac{KM_u}{N} - m$ with $0\leq \delta < 1$. Under the coded-caching strategy, the throughput (in bits) on the access links is 
$
		Q_{\rm cod, AC} = (1 - \delta)\frac{Q(K - m)}{m+1} + \delta \frac{Q(K-m-1)}{m + 2}$,
		and the backhaul thoughtput is 
$
		Q_{\rm cod, BH} = (1 - \delta)\left(1 - \left(\frac{M_b}{N}\right)^{m+1}\right)\frac{Q(K - m)}{m+1}  + \delta \left(1 - \left(\frac{M_b}{N}\right)^{m+2}\right)\frac{Q(K-m-1)}{m + 2}.
$
	\end{proposition}
	\begin{IEEEproof}
	We consider two cases: i) $M_u \in \{0, \frac{N}{K}, \frac{2N}{K}, \dots, \frac{(K-1)N}{K}\}$  and ii) $M_u$ has arbitrary value within $(0,N)$.
	
	\textbf{Case 1}: {$M_u \in \{0,\frac{N}{K}, \frac{2N}{K}, \dots, \frac{(K-1)N}{K}\}$} \\
	In this case, the user cache $M_u$ is multiple times of $\frac{N}{K}$. Denote $m = \frac{M_uK}{N} \in \{0, 1, 2, \dots, K-1\}$. When $m=0$, it is straightforward to see that $Q_{\rm cod,AC} = QK$ and $Q_{\rm cod,BH} = (1 - M_b/N)KQ$ since there is no cache at the users. The computation for $m\in \{1, \dots, K-1\}$ is as follows.
	
	\subsubsection*{Computation of $Q_{\rm cod, AC}$} We first calculate the total bits $Q_{\rm cod, AC}$ need to be sent over the access links under the coded-caching strategy. Let $\mathcal{CC}(\mathcal{F},\mathcal{K},m)$ denote the coded-caching algorithm that the BS employs to serve $K$ users. Each user is equipped with a cache of size $\frac{mN}{K}, m \in \mathbb{Z}^\star$, and requests one file from the library $\mathcal{F}$.  $\mathcal{CC}(\mathcal{F},\mathcal{K},m)$ comprises of two phases: a placement phase and a delivery phase. Due to space limitation, the details of $\mathcal{CC}(\mathcal{F},\mathcal{K},m)$ are omitted here but can be found in \cite[Sec.V]{AliNie2014}. We only present the essential information of $\mathcal{CC}(\mathcal{F},\mathcal{K},m)$ which will be used in the next subsection. Each file $F_f \in \mathcal{F}$ is divided into $\mathcal{C}^m_K$ non-overlapped subfiles. Then each file can be expressed as $F_f = (F_{f,\mathcal{T}}|\mathcal{T} \subset \mathcal{K}, |\mathcal{T}| = m)$, where $\mathcal{T}$ is any subset of $\mathcal{K}$ consisting of $m$ different elements. During the delivery phase, the BS multicasts $X_{\mathcal{S}} = \oplus_{s \in \mathcal{S}} F_{f_s, \mathcal{S}\backslash \{s\}}$ to all users, where $\mathcal{S} \subset \mathcal{K}$ with $|\mathcal{S}| = m+1$ and $\oplus$ denotes the XOR operation. It has been shown in \cite{AliNie2014} that 
	\begin{align}
	Q_{\rm cod, AC} = C^{m+1}_K \frac{Q}{C^m_K}
	= Q\frac{K - m}{1 + m}~(\text{bits}). \notag
	\end{align}
	
	\subsubsection*{Computation of $Q_{\rm cod, BH}$}
	Since the BS randomly stores parts of every file in its cache, the probability that a bit in file $F_f \in \mathcal{F}$ is prefetched at the BS cache is $p = \frac{M_b}{N}$. Now consider the transmission of signal $X_{\mathcal{S}}$. Each bit in $X_{\mathcal{S}}$ is the XORed of $m+1$ bits from $m+1$ different files. If these bits are available at the BS cache, there is no need to send this XORed bit through the backhaul. Otherwise, the data centre sends this XORed bit through the backhaul to the BS. Because these $m+1$ files are independent, the probability that this XORed bit is not sent through the backhaul is $p^{m+1}$. In other words, the probability that a XORed bit in $X_{\mathcal{S}}$ is sent through the backhaul is $1 - p^{m+1}$. Since there are $Q_{\rm cod,AC}$ XORed bits, the total bits sent through the backhaul is
	\begin{align*}
	Q_{\rm cod, BH} &= (1 - p^{m+1}) Q_{\rm cod,AC} \\
	&= \Big(1 - \Big(\frac{M_b}{N}\Big)^{m+1} \Big) \frac{Q(K-m)}{m+1}.
	\end{align*}
	
	\textbf{Case 2}: $0 < M_u < N$\\	
	This subsection calculates the throughput on the backhaul and access links for arbitrary values of the BS and user cache size. Let $m \in \mathbb{Z}^\star$ and $0 < \delta < 1$ such as $M_u = (m + \delta)\frac{N}{K}$. For every file $F_f \in \mathcal{F}$, we divide it into two parts: $F^1_f$ consisting of the first $(1-\delta)Q$ bits and $F^2_f$ consisting of the remaining $\delta Q$ bits. Then the original library $\mathcal{F}$ is decomposed into two disjoint sub-libraries $\mathcal{F}_1 = \{F^1_1, F^1_2, \dots, F^1_N\}$ and $\mathcal{F}_2 = \{F^2_1, F^2_2, \dots, F^2_N\}$. Note that the file size in $\mathcal{F}_1$ and $\mathcal{F}_2$ is $(1-\delta)Q$ and $\delta Q$ bits, respectively. 
	
	\subsubsection*{Cahe placement phase}
	The placement phase in this case comprises of two steps. First, the data centre applies the placement phase of $\mathcal{CC}(\mathcal{F}_1,\mathcal{K},m)$ on $\mathcal{F}_1$. After this step, each user cache contains $(1-\delta)M_uQ$ bits. Then, it applies $\mathcal{CC}(\mathcal{F}_2,\mathcal{K},m+1)$ on $\mathcal{F}_2$. This steps results in $\delta M_uQ$ bits on each user cache. In total, each user cache is prefetched with $(1 - \delta)M_uQ + \delta M_uQ = M_uQ$ bits, which satisfies the memory constraint. 
	
	\subsubsection*{Delivery phase}
	We employ a time-splitting mechanism to serve the user requests. As a result, the delivery phase consists of two consecutive steps. First, the delivery phase of $\mathcal{CC}(\mathcal{F}_1,\mathcal{K},m)$ is applied for $\mathcal{F}_1$. This will costs a throughput $(1 - \delta)\frac{Q(K - m)}{m +1}$ bits. Then the delivery phase of $\mathcal{CC}(\mathcal{F}_2,\mathcal{K},m+1)$ is applied for $\mathcal{F}_2$, which results in additional $\delta \frac{Q(K - m -1)}{m + 2}$ bits\footnote{This time-splitting mechanism can be seen as an implementation scheme to achieve the memory-sharing performance in \cite{AliNie2014}.}. Therefore, the total throughput on the access links is 
	\[
	Q_{\rm cod,AC} = (1 - \delta)Q\frac{K - m}{m +1} + \delta Q\frac{K - m - 1}{m + 2}.
	\]
	
	We observe that the probability that each XORed bit in $\mathcal{F}_1$ and $\mathcal{F}_2$ is stored at the BS cache is $q^{m+1} $ and $q^{m+2}$, respectively. Therefore, the backhaul throughput in this case is 
	\begin{align*}
	Q_{\rm cod,BH} =& ~(1 - \delta)\left(1 - q^{m+1} \right)Q\frac{K - m}{m +1}\notag \\
	&~ + \delta (1 - q^{m+2})Q\frac{K - m - 1}{m + 2}.
	\end{align*}
	\end{IEEEproof}
Proposition~\ref{prop:2} derives the aggregated throughput on the  access links under the coded-caching strategy for arbitrary values $M_u \in [0, N]$. When $\delta = 0$ and $\frac{KM_u}{N} \in \mathbb{Z}^\star$, the result is shorten as $\frac{KQ(1 - M_u/N)}{1 + KM_u/N}$, which can also be found in \cite{AliNie2014}. Note that \cite{AliNie2014} derives the access link's throughput only for limited values of $M_u$ such as $\frac{KM_u}{N} \in \mathbb{Z}$. In other words, Proposition~\ref{prop:2} generalizes the result in \cite{AliNie2014} for arbitrary values of the user cache size. 
	\subsection{Transmission model} \label{sec:transmission model}
	This subsection describes the transmission of the requested files from the BS to users.  Let $\mathbf{h}_{k} \in \mathbb{C}^{L\times 1}$ denote the channel vector from the BS antennas to user $k$, which follows  a circular-symmetric complex Gaussian distribution $\mathbf{h}_k \sim \mathcal{CN}(\mathbf{0}, \sigma^2_{h_k} \mathbf{I}_K)$, where $\sigma^2_{h_k}$ is the parameter accounting for the path loss from the BS antennas to user $k$. Perfect channel state information (CSI) is assumed to be available at the BS. In practice, robust channel estimation can be achieved through the transmission of pilot sequences. When a user requests a file, it first checks its own cache. If the requested file is available in its cache, it can be served immediately. Otherwise, the user sends the requested file's index to the data centre. If the requested file is not at the BS cache, it will be sent from the data centre via the backhaul. Then the BS transmits the requested file to the user via the access links. 
	
	\subsubsection{Signal transmission for uncoded caching strategy}
	 The data stream for each user under the uncoded caching method is transmitted independently. Denote $F_{d_1}, \dots, F_{d_K}$ as the requested files from user $1, \dots, K$, respectively, and $\bar{F}_{d_1}, \dots,\bar{F}_{d_K}$ as parts of the requested files which are not at the user cache. First, the BS modulates $\bar{F}_{d_k}$ in to corresponding modulated signal $x_k$ and then sends the precoded signal through the access channels. Denote $\mathbf{w}_k \in \mathbb{C}^{L \times 1}$ as the precoding vector for user $k$. The received signal at user $k$ is given as
	%
$
	y_k = \mathbf{h}^H_k \mathbf{w}_k x_k + \sum_{l\neq k} \mathbf{h}^H_k \mathbf{w}_l x_l + n_k, \label{eq:y_k}
	$
	%
	where $n_k$ is Gaussian noise with zero mean and variance $\sigma^2$. The first term in $y_k$ is the desired signal, and the second term is the inter-user interference. The signal-to-interference-plus-noise ratio at user $k$ is $
	\mathrm{SINR}_k = \frac{|\mathbf{h}^H_k \mathbf{w}_k|^2}{\sum_{l\neq k} |\mathbf{h}^H_k \mathbf{w}_l|^2 + \sigma^2}.
	$
	The information achievable rate of user $k$ is 
	\begin{align}
	R_{\mathrm{unc},k} = B\log_2\left(1 + \mathrm{SINR}_k \right), 1 \leq k \leq K, \label{eq:R_k}
	\end{align}
	where $B$ is the access links' bandwidth. 
	
	The transmit power on the access links under the uncoded caching policy is 
	$
	P_{\mathrm{unc}} = \sum_{k=1}^K \parallel\mathbf{w}_k\parallel^2.
	$

	\subsubsection{Signal transmission for coded caching strategy}
	Obviously, one can use the transmission design derived for the uncoded caching to delivery the requested files in the coded-caching method. However, since the coded caching strategy transmits a coded message to a group of users all users during the delivery phase, using the orthogonal beams might result in resources inefficiency. Thus, we employ physical-layer multicasting \cite{Sidirop2006} to precode the data for the coded caching strategy.

	In the coded caching strategy, the BS will send $C^{m+1}_K$ coded messages (of length $\frac{Q}{C^m_K}$ bits) in total to the users, each of which is received by a subset of $m+1$ users \cite{AliNie2014}. Denote by $\mathcal{S} \subset \mathcal{K}$ an arbitrary subset consisting of $m+1$ users, and by $\boldsymbol{\mathcal{S}} = \{\mathcal{S} ~|~ |\mathcal{S}| = m+1\}$ all subsets of $m+1$ users. Obviously, $|\boldsymbol{\mathcal{S}}| = C_K^{m+1}$. For convenience, we denote $X_{\mathcal{S}}$ as the coded message targeted for the users in $\mathcal{S}$. The received signal at user $k \in \mathcal{S}$ is given as
	%
$
	y_k = \mathbf{h}^H_k \mathbf{w}_{\mathcal{S}} x_{\mathcal{S}} + n_k$,
where $\mathbf{w}_{\mathcal{S}}$ is the beamforming vector for the users in $\mathcal{S}$ and $x_{\mathcal{S}}$ is the modulated signal of $X_{\mathcal{S}}$.
	The achievable rate for the users under physical-layer multicasting is
	\begin{align} \label{eq:r cod}
	R_{\mathrm{cod},\mathcal{S}} = \min_{k\in \mathcal{S}} \Big\{B\log_2\Big(1 + \frac{|\mathbf{h}^H_k \mathbf{w}_{\mathcal{S}}|^2}{\sigma^2}\Big)\Big\}.
	\end{align} 
	%
	The transmit power on the access links under the coded caching policy is 
	$
	P_{\rm{cod}} = \parallel\mathbf{w}_{\mathcal{S}}\parallel^2.
	$

	\section{Energy-efficiency analysis}\label{sec:EE}
	This section analyses the EE performance of the two caching strategies. 
	\begin{definition}[Energy efficiency]
		The EE measured in bit/Joule is defined as:
		\begin{align*}
		\mathrm{EE} = \frac{KQ}{E_{\Sigma}},
		\end{align*}
		where $KQ$ is the total requested bits from the $K$ users and $E_{\Sigma}$ is the total energy consumption for delivering these bits.
	\end{definition} 
	
	Since the cache placement phase in off-line caching occurs much less frequently (daily or weekly) than the delivery phase, we assume the energy consumption in the placement phase is negligible and thus $E_{\Sigma}$ is the energy cost in the delivery phase \cite{AliNie2014,TaoCheZhoYu2016}.
	\subsection{EE analysis for uncoded caching strategy}
	The total energy cost under the uncoded caching policy is given as $E_{\rm unc, \Sigma} = E_{\rm unc, BH} + E_{\rm unc, AC}$, where $E_{\rm unc, BH}$ and $E_{\rm unc, AC}$ are the energy cost on the backhaul and access links, respectively\footnote{In practice, $E_{\Sigma}$ also includes a static energy consumption factor.}. To compute the energy consumption on the access links, we note that each user requests $\frac{Q_{\rm unc, AC}}{K}$ bits. The uncoded caching strategy sends these bits to each user independently via unicasting. Since user $k$ requests a file at rate $R_{\mathrm{unc},k}$, it takes $\frac{Q_{\rm unc, AC}}{KR_{\mathrm{unc},k}}$ seconds to complete the transmission. Therefore, the total energy consumed on the access links is calculated as
		\begin{align}
		E_{\rm unc,AC} = \frac{Q_{\rm unc, AC}}{KR_{\mathrm{unc},k}} P_{\rm unc} = Q(1 - \frac{M_u}{N})\sum_{k=1}^K \frac{\parallel\mathbf{w}_k\parallel^2}{R_{\mathrm{unc},k}}. \notag
		\end{align}	

		Sine the backhaul link provides enough capacity to serve the access network, the energy cost on the backhaul is modelled as 
		\[
		E_{\rm unc,BH} = \eta Q_{\rm unc, BH} = \eta KQ\left(1- \frac{M_u}{N}\right)\left(1 - \frac{M_b}{N}\right),
		\]
		where $\eta$ is a constant. In practices, $\eta$ can be seen as the pricing factor used to trade energy for transferred bits \cite{TaoCheZhoYu2016}. The actual value of $\eta$ depends on the backhaul technology.

		Therefore, the EE under the uncoded caching strategy is given as
		\begin{align}
		\mathrm{EE}_{\rm unc} 
		= \frac{K}{ \left(1 - \frac{M_u}{N}\right)\left(\eta K\Big(1 -\frac{M_b}{N} \Big) 
			+ \sum_{k=1}^K\frac{ \parallel\mathbf{w}_k\parallel^2}{R_{\mathrm{unc},k}}\right)}.
		\label{eq:EE unc}
		\end{align}
		It is observed from \eqref{eq:EE unc} that $\mathrm{EE}_{\rm unc}$ is jointly determined by the cache capacities $M_u$ and $M_b$ and the transmitted power on the access links.
	\subsection{EE analysis for coded caching strategy}

	The energy cost on the backhaul link under the coded caching policy is given as $E_{\rm cod,BH} = \eta Q_{\rm cod,BH}$, where $\eta$ is the pricing factor. In order to calculate the energy consumption on the access links, $E_{\rm cod,AC}$, we note that the BS multicasts the coded information $X_{\mathcal{S}}$ to the users in $\mathcal{S}$. With the rate $R_{\rm cod,\mathcal{S}}$, it takes $\frac{Q_{\rm cod, AC}}{C^{m+1}_K R_{\mathrm{cod},\mathcal{S}}}$ seconds to send $X_{\mathcal{S}}$. The total energy consumed by the BS in this case is $E_{\rm cod, AC} = \frac{Q_{\rm cod, AC}}{C^{m+1}_K}\sum_{\mathcal{S} \in \boldsymbol{\mathcal{S}}} \frac{P_{\rm cod,\mathcal{S}}}{R_{\rm cod,\mathcal{S}}}$. Therefore, the EE under the coded caching strategy is given as
	\begin{align}
	\mathrm{EE}_{\rm cod} = \frac{KQ}{E_{\rm cod, \Sigma}} = \frac{KQ}{\eta Q_{\rm cod,BH} + \frac{Q_{\rm cod, AC}}{C^{m+1}_K}\sum_{\mathcal{S} \in \boldsymbol{\mathcal{S}}} \frac{P_{\rm cod,\mathcal{S}}}{R_{\rm cod,\mathcal{S}}}}. \label{eq:EE cod}
	\end{align}
	
	From Proposition~\ref{prop:2} we obtain 
	
$
	\mathrm{EE}_{\rm cod} = \frac{1 + \frac{KM_u}{N}}{\left(1 - \frac{M_u}{N}\right)\left( \eta \Big(1 - \left(\frac{M_b}{N}\right)^{\frac{KM_u}{N} + 1}\Big)  + 
		\frac{1}{C^{m+1}_K}\underset{\mathcal{S} \in \boldsymbol{\mathcal{S}}}{\sum} \frac{\parallel \mathbf{w}_{\mathcal{S}}\parallel^2}{R_{\rm cod,\mathcal{S}}}  \right) }. \label{eq:EE coded}
$
	
	Similarly, the EE under the coded-caching is determined by the BS and user storage capacity and the transmit power on the access links. 
	\subsection{Comparison between the two strategies}
	In general, the comparison between the two caching methods is complicated due to the contributions of many system parameters. In some cases, e.g., $\frac{KM_u}{N} \in \mathbb{Z}^\star$, however, it is possible to explicitly reveal each method's performance. Assuming that all users are served at the same rate, e.g., $R_{\mathrm{unc},k} = R_{\mathrm{cod},\mathcal{S}} = \gamma, \forall k, \mathcal{S}$.

	\subsubsection{Free-cost backhaul link}
	This occurs when the BS cache is large enough to store all the files, e.g., $M_b = N$ or $\eta = 0$. All the requested files are available at either user cache or BS cache. Consequently, we have:
	\[
	\mathrm{EE}_{\rm unc} = \frac{K}{\left(1 - \frac{M_u}{N}\right)\frac{P_{\rm unc}}{\gamma}}, ~~ \mathrm{EE}_{\rm cod} = \frac{1 + \frac{KM_u}{N}}{\left(1 - \frac{M_u}{N}\right)\frac{P_{\rm cod}}{\gamma}}.
	\]
When the two methods use the same transmit power on the access links, i.e., $P_{\rm unc} = P_{\rm cod}$, we have $\mathrm{EE}_{\rm unc} > \mathrm{EE}_{\rm cod}$. In general, the coded caching strategy will achieve a higher EE than the uncoded caching method when $M_u > \left(\frac{P_{\rm cod}}{P_{\rm unc}} - \frac{1}{K}\right)N$.
	\subsubsection{$M_b = 0$}
	In this case, all the requested files which are not at the user cache will be sent from the data centre, and thus
	\begin{align}
	\mathrm{EE}_{\rm unc}\! =\! \frac{1}{\left(1\! -\! \frac{M_u}{N} \!\right)\left(\!\eta \! +\! \frac{P_{\rm unc}}{\gamma K}\!\right) },\ 
	\mathrm{EE}_{\rm cod}\! =\! \frac{1 + \frac{KM_u}{N}}{\left(\!1\! -\! \frac{M_u}{N}\! \right) \left(\eta +\! \frac{P_{\rm cod}}{\gamma}\!\right)}. \nonumber
	\end{align}
	It is observed that the coded-caching strategy achieves higher EE than the uncoded caching method for the same transmit power since $\frac{KM_u}{N} > 0$ and $\frac{P_{\rm unc}}{K} < P_{\rm cod}$.

	\section{Energy-Efficiency Maximization in edge-caching wireless networks} \label{sec:EE maximization}
	We aim at maximizing the EE in edge-caching wireless networks under the two caching strategies. The general optimization problem is formulated as follows:
	\begin{align}
	\underset{\mathop \{\mathbf{w}_k\}_{k=1}^K, \mathbf{w}}{\mathtt{Maximize}} ~~ \mathrm{EE} \label{OP:EE} \qquad
	\mathtt{s.t.} ~~ \text{QoS constraint}, 
	\end{align}
	where $\mathrm{EE} \in \{\rm EE_{unc}, EE_{cod}\}$ and $R_k \in \{R_{\rm unc,k}, R_{\rm cod}\}$ which are given in Section~\ref{sec:transmission model}. 
	
\subsection{EE maximization for uncoded caching strategy}\label{sec:EE unc}
Let $\gamma_k$ denote the QoS requirement of user $k$ (bits per second). Without caching, it takes $t_k = \frac{Q}{\gamma_k}$ seconds to send user $k$ the requested file. However, since parts of the requested files are available in the user cache, the BS needs to send only $(1 - \frac{M_u}{N})Q$ bits to user $k$. Therefore, the rate requirement taking into account the user cache is $\bar{\gamma}_k = (1 - \frac{M_u}{N})Q/t_k = (1 - \frac{M_u}{N})\gamma_k$. It is observed from \eqref{eq:EE unc} that for a given network topology, the BS and user cache memories are fixed. Therefore, maximizing the EE is equivalent to minimizing the transmit power. Therefore, the problem \eqref{OP:EE} is equivalent to the following problem:
		%
		%
		%
		\begin{align}\label{OP:EE 1} 
		\underset{\mathop \{\mathbf{w}_k \in \mathbb{C}^L \}_{k=1}^K}{\mathtt{Minimize}} \sum_{k=1}^K\! \frac{\parallel\! \mathbf{w}_k\! \parallel^2}{R_{\mathrm{unc},k}}\!,~
		\mathtt{s.t.}~ \frac{|\mathbf{h}^H_k \mathbf{w}_k|^2}{\underset{l\neq k}{\sum}\! |\mathbf{h}^H_k \mathbf{w}_l|^2\! +\! \sigma^2} \geq \zeta_k, \forall k,
		\end{align}
		where the rate constraint is replaced by an equivalent SINR constraint $\zeta_k = 2^{\frac{\bar{\gamma}_k}{B}} - 1$.

	\subsubsection{Cost minimization by Zero-Forcing precoding}\label{sec:EE unc ZF}
	
	In this subsection, we maximize the EE based on the ZF design because of its low computational complexity. Since the direction of the beamforming vectors are already defined by the ZF, only transmitting power on each beam needs to be optimized. Let $p_k, 1\leq k \leq K,$ denote the transmit power dedicated for user $k$. The precoding vector for user $k$ is given as $\mathbf{w}_k = \sqrt{p_k} \tilde{\mathbf{h}}_k$, where $\tilde{\mathbf{h}}_k$ is the ZF beamforming vector for user $k$, which is the $k$-th column of  $\mathbf{H}^H (\mathbf{H}\mathbf{H}^H)^{-1}$, with $\mathbf{H} = [\mathbf{h}_1, \dots, \mathbf{h}_K]^T$. 
	\begin{theorem}
		Under the ZF design, the uncoded caching strategy achieves the maximal EE given as
		\begin{align}
		\mathrm{EE}^{\rm ZF}_{\rm unc} = \frac{K}{ \left(1 - \frac{M_u}{N}\right)\left(\eta K\Big(1 -\frac{M_b}{N} \Big) 
			+ \frac{ \sigma^2 \sum_{k=1}^K \zeta_k\parallel\tilde{\mathbf{h}}_k\parallel^2}{\bar{\gamma}_k}\right)}. \notag
		\end{align}
	\end{theorem}
	\begin{IEEEproof}
	By definition, $|\mathbf{h}^H_l \mathbf{w}_k|^2 = p_k\delta_{lk}$, where $\delta_{ij}$ is the Dirac delta function. Therefore, the constraint in \eqref{OP:EE 1} becomes $\frac{p_k}{\sigma^2} \geq \zeta_k,\forall k$. Consequently, the cost minimization problem is formulated as follows:
	\begin{align}\label{OP:EE unc 3}
	\underset{\mathop \{p_k: p_k \geq 0\}_{k=1}^K}{\mathtt{Minimize}} &~~ \sum_{k=1}^K \frac{a_kp_k}{\log_2(1 + a_kp_k/\sigma^2 )} \\
	\mathtt{s.t.} &~~  p_k \geq \zeta_k
	\sigma^2, \forall k,  \notag
	\end{align}
	where $a_k = \parallel\! \tilde{\mathbf{h}}_k\!\parallel^2$.
	
	Consider a function $f(x) = \frac{ax}{\log_2(1 + bx)}$ with $a,b \geq 0$ in $\mathbb{R}^+$. The derivative of $f(x)$ is $f'(x) = \frac{a}{\log_2(1 + bx)}\Big(1 - \frac{bx}{\log(1 + bx)(1+bx)}\Big) > 0, \forall x > 0$. Therefore, the objective function of \eqref{OP:EE unc 3} is a strictly increasing function in its supports. Therefore, the optimal solution of \eqref{OP:EE unc 3} is achieved at 
	$
	p^\star_k = \zeta_k \sigma^2,
	$
	and the minimum transmit power is $ \sigma^2\sum_{k=1}^K \zeta_k \parallel\!\tilde{\mathbf{h}}_k\!\parallel^2$. Substituting this into $\mathrm{EE}_{\rm unc}$, we obtain the proof of Theorem 1. 
\end{IEEEproof}

	\subsubsection{Cost minimization by Semi-Definite Relaxation} \label{sec:EE unc SDR}
	In this subsection, we maximize the EE by design the beamforming vectors and power allocation simultaneously. It is seen that \eqref{OP:EE 1} is a NP-hard problem due to its non-convex objective functions as well as the constraints. Therefore, we resort to solve a suboptimal solution of \eqref{OP:EE 1} by minimizing the upper bound of the objective function. Since it requires $R_{\mathrm{unc},k} \geq \bar{\gamma}_k$ to deliver the requested content to the users successfully, we have $ \frac{\parallel\! \mathbf{w}_k\! \parallel^2}{R_{\mathrm{unc},k}} \leq \frac{\parallel\! \mathbf{w}_k\! \parallel^2}{\bar{\gamma}_k}$. Due to the difference of transmission time among the users, a user who has received the requested file may not interfere the transmission of other users. Denote $\mathcal{K}_t \triangleq \{k~|~\bar{\gamma}_k \leq \frac{Q}{t}, \forall t \in [0, \frac{Q}{\min_k (\bar{\gamma}_k)}]\}$ as the subset of active users at the time of interest. Then the resorted problem is stated as
	\begin{align}\label{OP:EE 11} 
	\underset{\mathop \mathbf{w}_k \in \mathbb{C}^L }{\mathtt{Minimize}} \sum_{k\in \mathcal{K}_t}\! \frac{\parallel\! \mathbf{w}_k\! \parallel^2}{\bar{\gamma}_k}\!,~
	\mathtt{s.t.}~ \frac{|\mathbf{h}^H_k \mathbf{w}_k|^2}{\underset{k\neq l \in \mathcal{K}_t}{\sum}\! |\mathbf{h}^H_k \mathbf{w}_l|^2\! +\! \sigma^2} \geq \zeta_k, \forall k.
	\end{align}
	%
	
	We introduce new variables $\mathbf{X}_k = \mathbf{w}_k \mathbf{w}^H_k \in \mathbb{C}^{L\times L}$ and denote $\mathbf{A}_k = \mathbf{h}_k \mathbf{h}^H_k \in \mathbb{C}^{L\times L}$. Since $|\mathbf{h}^H_l \mathbf{w}_k|^2 = \mathbf{h}^H_l \mathbf{w}_k \mathbf{w}^H_k \mathbf{h}_l = \mathrm{Tr}(\mathbf{h}_l \mathbf{h}^H_l \mathbf{w}_k \mathbf{w}^H_k) = \mathrm{Tr}(\mathbf{A}_l \mathbf{X}_k)$, we can reformulate problem \eqref{OP:EE 11} as
	\begin{align}
	\underset{\mathop \mathbf{X}_k \in \mathbb{C}^{L \times L}}{\mathtt{Minimize}}& ~~ \sum_{k\in \mathcal{K}_t} \mathrm{Tr}(\mathbf{X}_k) \label{OP:EE unc 2}\\ 
	\mathtt{s.t.}& ~~\!  \mathrm{Tr}(\!\mathbf{A}_k \mathbf{X}_k\!) \geq \zeta_k\! \sum_{k\neq l \in \mathcal{K}_t}\! \mathrm{Tr}(\mathbf{A}_l \mathbf{X}_k)\! +\! \zeta_k \sigma^2, \forall k, \notag\\
	&~~ \mathbf{X}_k \succeq \mathbf{0}, \mathrm{rank}(\mathbf{X}_k) = 1, \forall k. \notag
	\end{align}
	Problem \eqref{OP:EE unc 2} is still difficult to solve because the rank-one constraint is non-convex. Fortunately, the objective function and the two first constraits are convex. Therefore, \eqref{OP:EE unc 2} can be effectively solved by the SDR which is obtained by ignoring the rank one constraint. Since the SDR of \eqref{OP:EE unc 2} is a convex optimization problem, it can be effectively solved by using, e.g., the primal-dual interior point method \cite{Boyd2004}. Gaussian randomization procedure may be used to compensate the ignorance of the rank-one constraint in the SDR solution \cite{Luo2010}. It has been shown that SDR can achieve a performance close to the optimal solution \cite{Luo2010}.  From the solution $\mathbf{X}^\star_k$ of the SDR of \eqref{OP:EE unc 2}, we obtain the precoding vector $\mathbf{w}^\star_k$. Substituting $\mathbf{w}^\star_k$ into (\ref{eq:EE unc}) we obtain the EE of the uncoded caching strategy under SDR design.

	\subsection{EE maximization for coded caching strategy}
	Given the QoS requirement $\gamma_k$, user $k$ expects to receive the requested file in $t_k = \frac{Q}{\gamma_k}$. Since each user receives only $C^m_{K-1}$ coded messages out of $C^{m+1}_K$, the active time for user $k$ is $\frac{C^m_{K-1}}{C^{m+1}_K} t_k = \frac{(m+1)Q}{K\gamma_k}$. Therefore, the required rate for user $k$ is $\bar{\gamma}_k = (\frac{Q*C^m_{K-1}}{C^m_K})/(\frac{(m+1)Q}{K\gamma_k}) = \frac{K-m}{m+1}\gamma_k$, where $\frac{Q*C^m_{K-1}}{C^m_K}$ is the number of coded bits sent to user $k$. 
	Since the cache memories \eqref{eq:EE cod} are constant, maximizing the EE is equivalent to minimizing  $\frac{P_{\rm cod}}{R_{\mathrm{cod},\mathcal{S}}}$, where $R_{\mathrm{cod},\mathcal{S}}$ is given in (\ref{eq:r cod}). The optimization problem in this case is stated as
	\begin{align}
	\underset{\mathop \mathbf{w}_{\mathcal{S}}\in \mathbb{C}^{L\times 1}}{\mathtt{Minimize}}& ~~ \frac{\parallel\! \mathbf{w}_{\mathcal{S}}\! \parallel^2}{R_{\mathrm{cod},\mathcal{S}}}, \quad 	\mathtt{s.t.} ~ R_{\mathrm{cod},\mathcal{S}} \geq \bar{\gamma}_k, \forall k\in \mathcal{S}.  \label{OP:EE cod}
	\end{align}
	We note that problem \eqref{OP:EE cod} optimizes the beamforming vector for only a subset of users in $\mathcal{S}$. Because $\frac{P_{\rm cod}}{R_{\mathrm{cod},\mathcal{S}}}$ is not convex, we instead find a suboptimal solution of problem \eqref{OP:EE cod} by minimizing the upper bound of  $\frac{P_{\rm cod}}{R_{\mathrm{cod},\mathcal{S}}}$, i.e.,  $\frac{P_{\rm cod}}{R_{\mathrm{cod},\mathcal{S}}} \leq  \frac{P_{\rm cod}}{\bar{\gamma}_{\rm min, \mathcal{S}}}$, where $\bar{\gamma}_{\rm min,\mathcal{S}} = \min_{k\in \mathcal{S}} \bar{\gamma}_k$. 
	By introducing a new variable $\mathbf{X} = \mathbf{w}_{\mathcal{S}}^H \mathbf{w}_{\mathcal{S}} \in \mathbb{C}^{L\times L}$, the reformulated problem is given as 
	\begin{align}
	\underset{\mathop \mathbf{X} \in \mathbb{C}^{L \times L}}{\mathtt{Minimize}} \frac{\mathrm{Tr}(\mathbf{X})}{\bar{\gamma}_{\rm min,\mathcal{S}}}, \label{OP:EE cod 1} 
	&~~\mathtt{s.t.} ~\! \mathbf{X} \succeq \mathbf{0}; ~ \mathrm{rank}(\mathbf{X}) = 1; \\
	&~~  \mathrm{Tr}(\!\mathbf{A}_k \mathbf{X}\!)\! \geq\!  \sigma^2 (2^{\frac{\bar{\gamma}_{\rm min,\mathcal{S}}}{B}}\! -\! 1), \forall k \in \mathcal{S}. \notag
	\end{align}
	%
	We observe that the objective function and the constraints of problem \eqref{OP:EE cod 1} are convex, except the rank-one constraint. This suggests to solve problem \eqref{OP:EE cod 1} via SDR method by ignoring the rank-one constraint. It is noted that the solution of SDR does not always satisfy the rank-one condition. Thus, Gaussian randomization procedure might be used to obtain the approximated vector from the SDR solution \cite{Luo2010}. From the solution $\mathbf{X}^\star$ of problem \eqref{OP:EE cod 1}, we obtain the precoding vector $\mathbf{w}^\star_\mathcal{S}$. Substituting $\mathbf{w}^\star_\mathcal{S}$ into \eqref{eq:EE cod}, we obtain the EE for the coding caching strategy.
	%
	%
	%
	\section{Minimization of content delivery time}\label{sec:delivery time}
	In this section, we aim at minimizing the average time for delivering the requested files to all users. In general, the delivery time is comprised of two parts caused by the backhaul and access links. In practice, the backhaul capacity is usually much greater than the access capacity. Therefore, we assume negligible delivery time on the backhaul link. It is also assumed that the processing time at the BS is fixed and negligible. Therefore, the total delivery time is mainly determined by the access links.

\subsection{Minimization of delivery time for uncoded caching strategy}\label{sec:time unc}
We would like to remind here that the uncoded caching strategy transmits independent data streams to the users. Let $t_k$ be a time duration for the BS to transmit all the $\frac{Q_{\rm unc,AC}}{K}$ requested bits to user $k$. Since the BS serves user $k$ with rate $R_{\mathrm{unc},k}$, we have $t_k = \frac{Q_{\rm unc,AC}}{K R_{\mathrm{unc},k}}$ seconds. The average delivery time in the uncoded caching strategy is given as 
\[
\tau_{\rm unc} = \frac{1}{K}\sum_{k=1}^K t_k = \frac{Q(1 - \frac{M_u}{N})}{K}\sum_{k=1}^K \frac{1}{R_{\mathrm{unc},k}}.
\]
The minimization of $\tau_{\rm unc}$ is formulated as
\begin{align}
\underset{\mathop \{\mathbf{w}_k \in \mathbb{C}^L\}_{k=1}^K}{\mathtt{Minimize}}& ~~ \frac{Q(1 - \frac{M_u}{N})}{K}\sum_{k=1}^K \frac{1}{R_{\mathrm{unc},k}} \label{OP:time unc} \\
\mathtt{s.t.}&~~ R_{\mathrm{unc},k} \geq \bar{\gamma}_k, \forall k; ~ \sum_{k=1}^K \parallel\! \mathbf{w}_k\! \parallel^2 \leq P_{\Sigma}, \notag
\end{align}

where the first constraint is to satisfy the QoS requirement and $P_{\Sigma}$ is the total transmit power. 

\subsubsection{Zero-Forcing precoding design}
Let $\tilde{\mathbf{h}}_k$ be the ZF precoding vector for user $k$, which is the $k$-th column of the ZF precoding matrix $\mathbf{H}^H (\mathbf{H}\mathbf{H}^H)^{-1}$. The beamforming vector is parallel to the ZF precoding vector as $\mathbf{w}_k = \sqrt{p_k} \tilde{\mathbf{h}}_k$, where $p_k$ is the power allocating to user $k$. Note that under the ZF precoding, $\mathbf{h}^H_l \tilde{\mathbf{h}}_k = \delta_{lk}$, we thus have $R^{ZF}_{\mathrm{unc},k} = \log_2 (1 + \frac{p_k}{\sigma^2})$. Therefore, the problem \eqref{OP:time unc} is equivalent to 
\begin{align} \label{OP:time min ZF}
\underset{\mathop \{p_k: p_k \geq 0\}_{k=1}^K}{\mathtt{Minimize}}& ~~ \frac{Q(1 - \frac{M_u}{N})}{K}\sum_{k=1}^K \frac{1}{\log_2(1 + p_k/\sigma^2)} \\
\mathtt{s.t.}&~~ \frac{p_k}{\sigma^2} \geq \zeta_k, \forall k; ~ \sum_k p_k\!\parallel\! \tilde{\mathbf{h}}_k\! \parallel^2 \leq P_{\Sigma}. \notag
\end{align}
\begin{proposition}
	Given the total power $P_{\Sigma}$ satisfying $P_{\Sigma} \geq  \sigma^2 \sum_{k=1}^K\zeta_k\parallel\! \tilde{\mathbf{h}}_k\! \parallel^2$, the problem \eqref{OP:time min ZF} is convex and feasible.
\end{proposition}
\begin{IEEEproof}
We will show that $P_{\Sigma} \geq  \sigma^2 \sum_{k=1}^K \zeta_k \parallel\! \tilde{\mathbf{h}}_k\! \parallel^2$ is the necessary and sufficient conditions of problem \eqref{OP:time min ZF}. It is straightforward to see that the constraints of \eqref{OP:time min ZF} are convex. We will show that the objective function is also convex. Indeed, consider the function $f(x) = 1/\log_2(1 + ax)$ in $\mathbb{R}^+$ with $a > 0$. The second-order derivative of $f(x)$ is given as 
\begin{align*}
	f'(x) &= -\frac{a}{\log_2(1 + ax) (1 + ax)},\\
	f''(x) &=  \frac{a^2}{\log^2_2(1 + ax) (1+ax)^2} + 
	\frac{a^2}{\log_2(1+ax)(1 + ax)^2}.
\end{align*}
It is verified that the second-order derivative is always positive, thus the objective function is convex in its support. Consequently, this problem can effectively solved by efficient algorithms, e.g., CVX \cite{Boyd2004}.

Now assuming that the problem \eqref{OP:time min ZF} is feasible. Then there exists a solution $\{\bar{p}_k\}_{k=1}^K$ which satisfies all the constraints. From the first constraint, it is straightforward to verify that $P_{\Sigma} \geq \sigma^2 \sum_{k=1}^K \zeta_k\parallel\! \tilde{\mathbf{h}}_k\! \parallel^2$.
\end{IEEEproof}
\subsubsection{General beamforming design}
Finding the optimal solution of the original problem \eqref{OP:time unc} is challenging because of the non-convex objective function. We instead propose to solve \eqref{OP:time unc} sub-optimally via minimizing the upper bound of $\tau_{\rm unc}$. Since 
\[
\tau_{\rm unc} \leq \max \{t_1, \dots, t_K\} = \frac{Q(1-\frac{M_u}{N})}{\min \{R_{\mathrm{unc},1}, \dots, R_{\mathrm{unc},K}\}},
\]
and $Q(1 - \frac{M_u}{N})$ is a positive constant, the suboptimal optimization of \eqref{OP:time unc} is formulated as
\begin{align}
\underset{\mathop \{\mathbf{w}_k \in \mathbb{C}^L \}_{k=1}^K}{\mathtt{Maximize}}& ~~ \min \{R_{\mathrm{unc},1}, \dots, R_{\mathrm{unc},K}\} \label{OP:max min code}\\
\mathtt{s.t.}&~~ R_{\mathrm{unc}, k} \geq \bar{\gamma}_k, \forall k; ~ \sum_k \parallel\! \mathbf{w}_k\! \parallel^2 \leq P_{\Sigma}. \notag
\end{align}
By introducing an arbitrary positive variable $x$ and resorting to SINR constraint, the above problem is equivalent to
\begin{align}
\underset{\mathop x>0, \{\mathbf{w}_k\in \mathbb{C}^L \}_{k=1}^K}{\mathtt{Maximize}} ~ x, \label{OP:time unc 1} ~
\mathtt{s.t.}&~ \frac{|\mathbf{h}^H_k \mathbf{w}_k|^2}{\sum_{l\neq k} |\mathbf{h}^H_k \mathbf{w}_l|^2\! +\! \sigma^2} \geq x, \forall k, \\
&~ x \geq \zeta_k;~ \sum_k \parallel\! \mathbf{w}_k\! \parallel^2 \leq P_{\Sigma}, \notag
\end{align}
%

%

We introduce new variables $\mathbf{X}_k = \mathbf{w}_k \mathbf{w}^H_k$ and remind that $\mathbf{A}_k = \mathbf{h}_k \mathbf{h}^H_k$. The problem \eqref{OP:time unc 1} is equivalent to
\begin{align}\label{OP:24}
\underset{\mathop \{\mathbf{X}_k \in \mathbb{C}^{L\times L}\}_{k=1}^K, x}{\mathtt{Maximize}} ~~ &x, ~~~ 
	\mathtt{s.t.} ~ x \geq \zeta_k;~ \sum_k \mathrm{Tr}(\mathbf{X}_k) \leq P_{\Sigma}; \\
& \mathrm{Tr}(\mathbf{A}_k\mathbf{X}_k) - x \sum_{l\neq k} \mathrm{Tr}(\mathbf{A}_k \mathbf{X}_l) \geq x\sigma^2, \forall k;\notag \\
& \mathbf{X}_k \succeq \mathbf{0};~~ \mathrm{rank}(\mathbf{X}_k) = 1. \notag
\end{align}
\begin{table}
	\centering
	\caption{\textsc{Algorithm to solve \eqref{OP:24}}}\label{table:1}
	\begin{tabular}{l l}
		\hline
		\vspace{-0.0in} & \\
		1. & Initialize $A_H$, $A_L = \zeta$, and the accuracy $\epsilon$. \\
		2. & $A_M$ = $(A_H+A_L)/2$. \\
		3. & Given $A_M$, if (\ref{OP:25}) is feasible, then $A_L := A_M$. \\
		& Otherwise $A_H := A_M$. \\
		4. & Repeat step 2 and 3 until $|A_H-A_L| \leq \epsilon$. \\
		\hline
	\end{tabular} 
	\vspace{-0.0cm}
\end{table}
It is observed that the third constraint is convex for a given $x$. Therefore, the SDR solution of problem \eqref{OP:24}, which is obtained by ignoring the rank one constraint, can be solved via bisection. The steps to solve are given in Table~\ref{table:1}.
\begin{align} \label{OP:25}
\mathtt{find}&~~ \{\mathbf{X}_k \in \mathbb{C}^{L\times L}\}_{k=1}^K \\
\mathtt{s.t.}&~~ \mathrm{Tr}(\mathbf{A}_k\mathbf{X}_k) - A_M \Big(\sum_{l\neq k} \mathrm{Tr}(\mathbf{A}_k \mathbf{X}_l) +\sigma^2\Big) \geq 0, \forall k\notag \\
&~~ \sum_k \mathrm{Tr}(\mathbf{X}_k) \leq P_{\Sigma}; ~~ \mathbf{X}_k \succeq \mathbf{0}, \forall k. \notag
\end{align}
%
%
%

\subsection{Minimization of delivery time for coded caching strategy}
The coded caching strategy multicasts the coded message $\mathbf{X}_{\mathcal{S}}$ to the users in $\mathcal{S}$. Since each $\mathbf{X}_{\mathcal{S}}$ contains $\frac{Q_{\rm cod, AC}}{C^{m+1}_K}$ bits, the delivery time under coded-caching strategy is 
$\tau_{\rm cod} = \frac{Q_{\rm cod, AC}}{C^{m+1}_K} \sum_{\mathcal{S}\in \boldsymbol{\mathcal{S}}} \frac{1}{R_{\rm cod,\mathcal{S}}}$, where $R_{\rm cod,\mathcal{S}}$ is given in \eqref{eq:r cod}. Since the transmissions of $X_{\mathcal{S}}$ are independent, the optimization problem of $\tau_{cod}$ becomes minimizing the delivery time of each $X_{\mathcal{S}}$, as follows:
\begin{align}
\underset{\mathop \mathbf{w}_{\mathcal{S}} \in \mathbb{C}^L}{\mathtt{Minimize}}& ~~   \frac{1}{R_{\rm cod,\mathcal{S}}} \\
\mathtt{s.t.}&~~ R_{\mathrm{cod},\mathcal{S}} \geq \bar{\gamma}_{\rm min,\mathcal{S}};~ \parallel\! \mathbf{w}_{\mathcal{S}}\! \parallel^2 \leq P_{\Sigma}. \notag
\end{align}
By introducing new variables $x > 0$, $\mathbf{X} = \mathbf{w}_{\mathcal{S}} \mathbf{w}^H_{\mathcal{S}} \in \mathbb{C}^{L\times L}$ and using the equivalent SINR constraint, the above optimization is equivalent to
\begin{align}\label{OP:time cod}
\underset{\mathop x, \mathbf{X} \in \mathbb{C}^{L\times L}}{\mathtt{Maximize}} &~~ x \\
\mathtt{s.t.}&~~ \mathrm{Tr}(\mathbf{A}_k\mathbf{X}) \geq x\sigma^2, \forall k \in \mathcal{S};~\mathbf{X} \succeq \mathbf{0};  \notag\\
&~~ x \geq 2^{\bar{\gamma}_{\rm min,\mathcal{S}}} - 1;~ \mathrm{Tr}(\mathbf{X}) \leq  P_{\Sigma};~ \mathrm{rank}(\mathbf{X}) = 1.\notag
\end{align}
\begin{table}
	\centering
	\caption{\textsc{Algorithm to solve \eqref{OP:time cod}}}\label{table:2}
	\begin{tabular}{l l}
		\hline
		\vspace{-0.0in} & \\
		1. & Initialize $A_H$, $A_L = 2^{\bar{\gamma}_{\rm min,\mathcal{S}}} - 1$, and the accuracy $\epsilon$. \\
		2. & $A_M$ = $(A_H+A_L)/2$. \\
		3. & Given $A_M$, if (\ref{OP:time cod sdr}) is feasible, then $A_L := A_M$. \\
		& Otherwise $A_H := A_M$. \\
		4. & Repeat step 2 and 3 until $|A_H-A_L| \leq \epsilon$. \\
		\hline
	\end{tabular} 
	\vspace{-0.3cm}
\end{table}
Similar to the previous subsection, we observe that the first constraint in \eqref{OP:time cod} is convex for a given $x$. Therefore, the above optimization can be solved via bisection and SDR by removing the rank one constraint. The steps to solve are given in Table~\ref{table:2}.
\begin{align}
\mathtt{find}&~~ \mathbf{X} \in \mathbb{C}^{L\times L}  \label{OP:time cod sdr} \\
\mathtt{s.t.}&~~ \mathrm{Tr}(\mathbf{A}_k\mathbf{X}) - A_M \sigma^2 \geq 0, \forall k\in \mathcal{S} \notag \\
&~~ \mathrm{Tr}(\mathbf{X}) \leq P_{\Sigma};~~ \mathbf{X} \succeq \mathbf{0}. \notag
\end{align}
\section{Non-uniform file popularity distribution}\label{sec:Zipf}
In most practical cases, the content popularity does not follow uniform distribution. In fact, there are always some files which are more frequently requested than the others. In this section, we consider arbitrary user content popularity and the uncoded caching strategy. Let $\mathbf{p}_k = \{q_{k,1}, \dots, q_{k,N}\}$ with $\sum_{n=1}^N q_{k,n} = 1$ denote the content popularity of user $k$, where $q_{k,n}$ is the probability of the $n$-th file being requested from user $k$. 

The global file population at the BS is computed as follows:
\begin{align}
	q_{G,n} = \frac{1}{K}\sum_{k=1}^K q_{k,n}.
\end{align}

We consider general cache memories in which the user caches' size can be different. For convenience, let $M_0$ (files) denote the storage memory at the BS and $M_k$ (files) denote the storage memory at user $k$. In the placement phase, each user fills its cache based on the local file popularity until full. Denote $\tilde{\mathbf{q}}_k = \Pi(\mathbf{q}_k)$ and $\tilde{\mathbf{q}}_G = \Pi(\mathbf{q}_G)$ as the sorted version in decreasing order of $\mathbf{q}_k$ and $\mathbf{q}_G$, respectively. Then user $k$ stores the first $n_k = M_k$ files in $\tilde{\mathbf{q}}_k$. Similarly, the BS stores the first $n_G = M_0$ files in $\tilde{\mathbf{q}}_G$.

In the delivery phase, the users send their requested file indices to the data centre. 
	\begin{proposition}\label{prop:3}
	Let $\mathbf{D} = \{d_1, \dots, d_K\}$ denote a set of file indices which are requested by the users. The total throughput on the access links is given as
	\begin{align}
		Q_{\rm AC}(\mathbf{D}) = Q\sum_{k=1}^K \mathbb{I}_{n_k}(\Pi_k(d_k))
	\end{align}
and the backhaul's throughput is calculated as
\begin{align}
	Q_{\rm BH}(\mathbf{D}) = Q\sum_{k=1}^K \mathbb{I}_{n_G}(\Pi_G(d_k)),
\end{align}
where $\Pi_k(d_k)$ is the new position of file $d_k$ after sorted by $\Pi(\mathbf{q}_k)$, and  $\mathbb{I}_n(i) = 1$ if $i>n$ and $0$ otherwise.
\end{proposition}
The proof of Proposition~\ref{prop:3} is straightforward followed by checking if the requested file is available at the BS or user caches.

In this caching strategy, a user stores the whole file if it is cached. Therefore, the BS will transmit only to a subset of users $\tilde{\mathcal{K}}(\mathbf{D}) = \{k~|~\Pi_k(d_k) > n_k\}$ who do not cache the requested files. In order to minimize the energy cost, the BS applies the signal transmission design as follows:
\begin{align}\label{OP:nonuniform EE} 
	\underset{\mathop \mathbf{w}_{k\in \tilde{\mathcal{K}}(\mathbf{D})} \in \mathbb{C}^L }{\mathtt{Minimize}}~~&~~ \sum_{k \in \tilde{\mathcal{K}}(\mathbf{D})} \frac{\parallel \mathbf{w}_k\! \parallel^2}{\tilde{R}_{\mathrm{unc},k}}, \\
	\mathtt{s.t.}&~~ \tilde{R}_{\mathrm{unc},k} \geq \gamma, \forall k \in \tilde{\mathcal{K}}(\mathbf{D}),\notag
\end{align}
where $\tilde{R}_{\mathrm{unc},k} = B\log_2\Big(1 + \frac{|\mathbf{h}^H_k \mathbf{w}_k|^2}{\sum_{k\neq l \in \tilde{\mathcal{K}}(\mathbf{D})} |\mathbf{h}^H_k \mathbf{w}_l|^2 + \sigma^2}\Big)$.

The delivery time minimization problem is formulated as:
\begin{align}\label{OP:nonuniform time} 
	\underset{\mathop \mathbf{w}_{k\in \tilde{\mathcal{K}}(\mathbf{D})} \in \mathbb{C}^L }{\mathtt{Minimize}}~~&~~ \sum_{k \in \tilde{\mathcal{K}}(\mathbf{D})} \frac{Q}{\tilde{R}_{\mathrm{unc},k}}, \\
	\mathtt{s.t.}&~~ \tilde{R}_{\mathrm{unc},k} \geq \gamma, \forall k \in \tilde{\mathcal{K}}(\mathbf{D}).\notag
\end{align}

The solution of problem \eqref{OP:nonuniform EE} and \eqref{OP:nonuniform time} can be found by similar techniques in Section~\ref{sec:EE unc} and Section~\ref{sec:time unc}, respectively.

\begin{table}
	\centering
	\caption{Simulation time in seconds, $m=K-1$}
	\begin{tabular}{ |p{0.6cm}|p{1.6cm}|p{1.6cm}|p{1.6cm}|  }
		\hline
		K & Coded & Uncoded-SDR & Uncoded-ZF\\
		\hline
		4 & 0.197 & 0.384 & 8.7e-5\\
		8 & 0.204 & 1.131 & 10e-5\\
		\hline
	\end{tabular}\label{table:3}
\end{table}
\begin{figure*}
	\normalsize 
	\begin{center}
	\subfigure[Cost-free backhaul]{\includegraphics[width= \columnwidth]{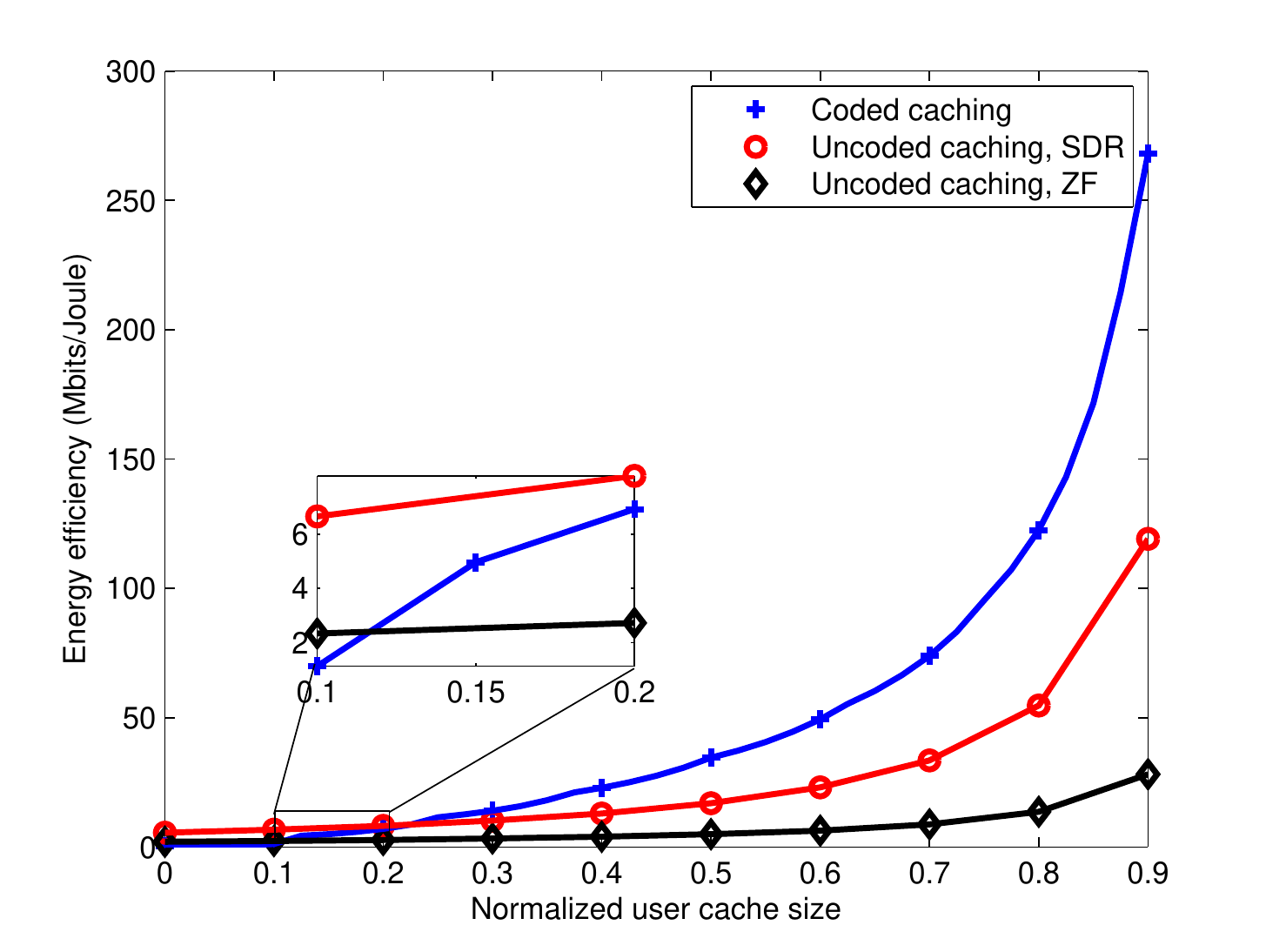}}	
	\subfigure[$M_b = 0.7N$] {\includegraphics[width=\columnwidth]{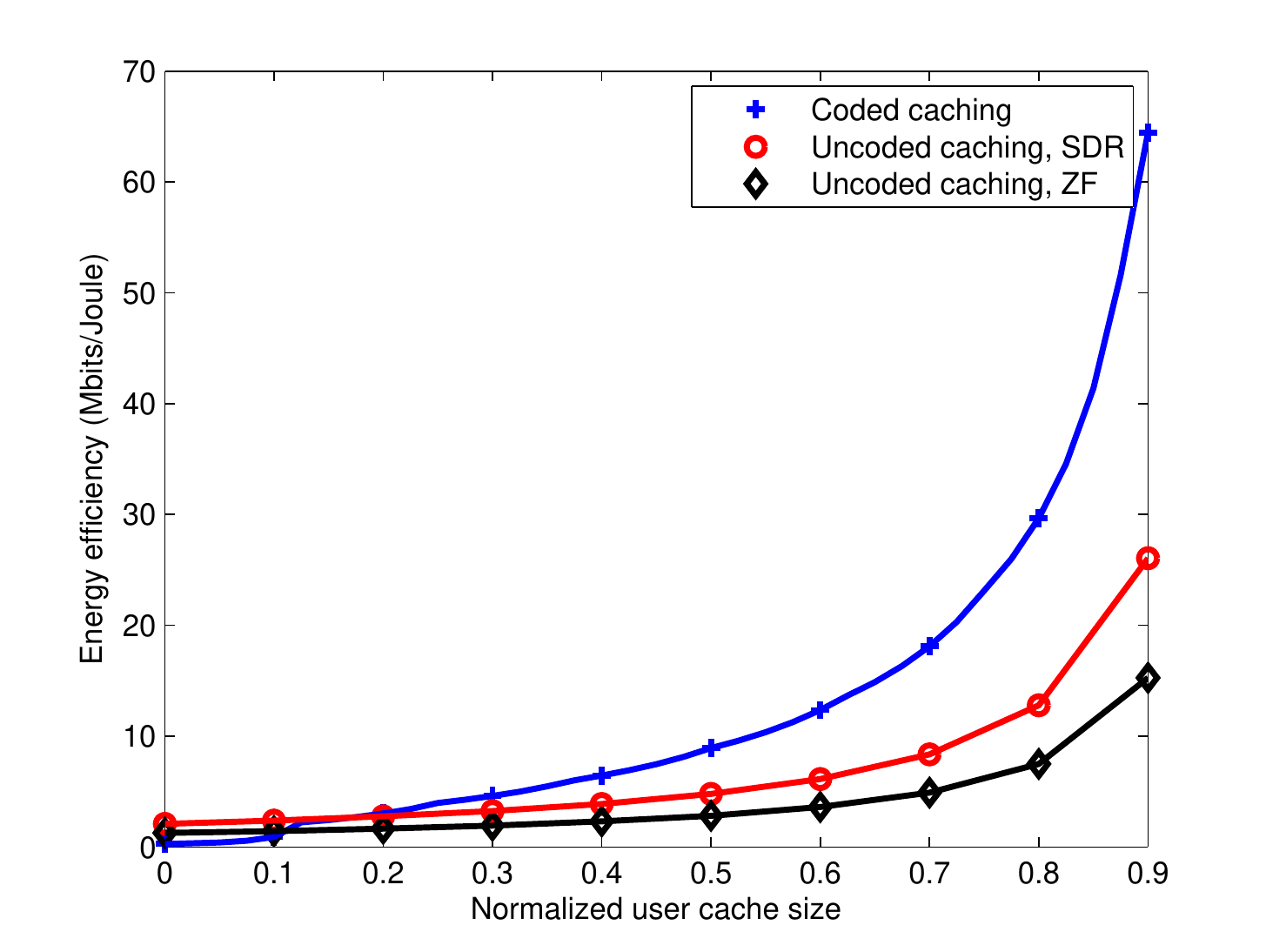}}
	\subfigure[$M_b = 0.1N$] {\includegraphics[width=\columnwidth]{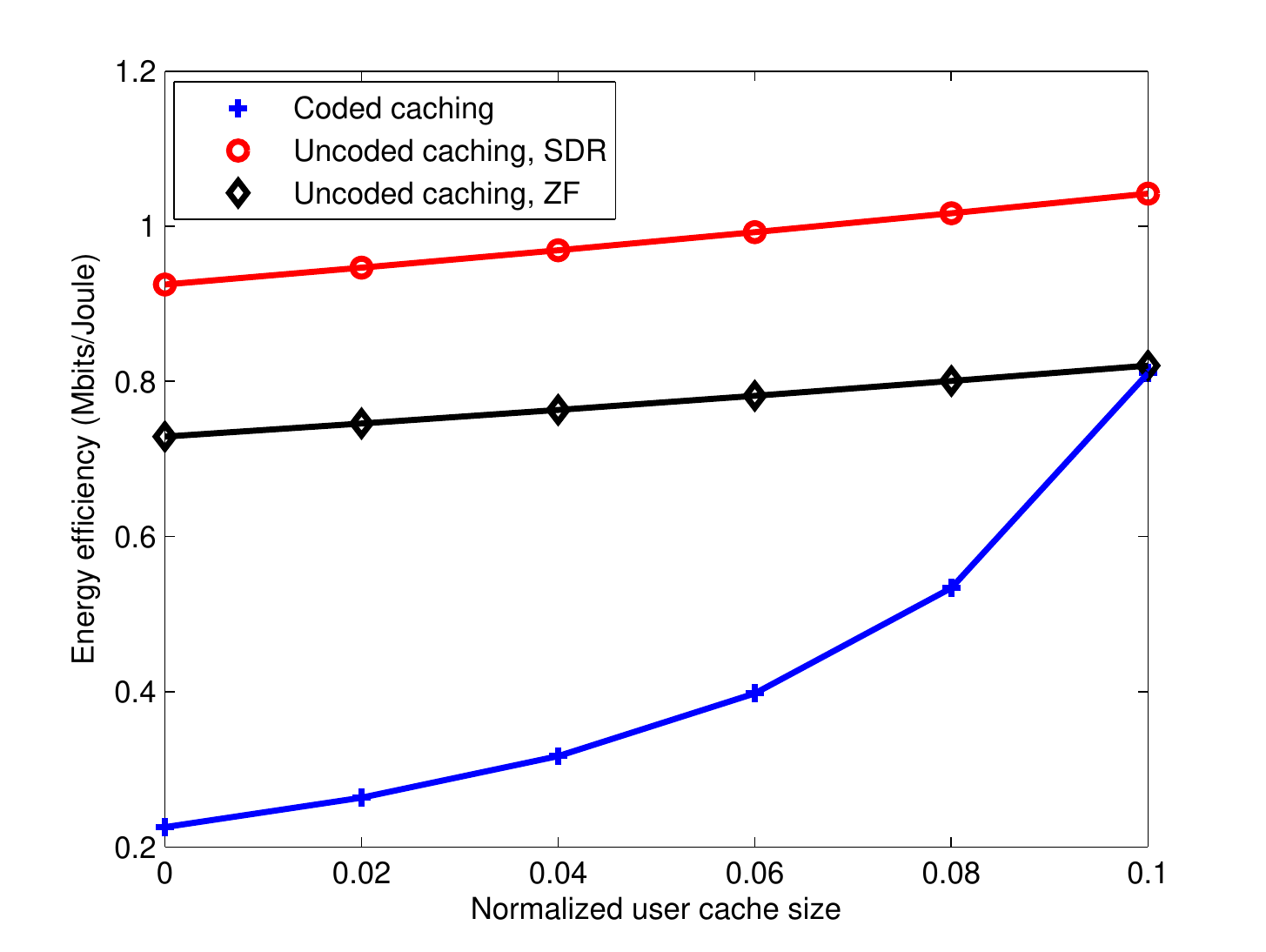}}	
	\subfigure[$M_u = 0.5N$]{\includegraphics[width=\columnwidth]{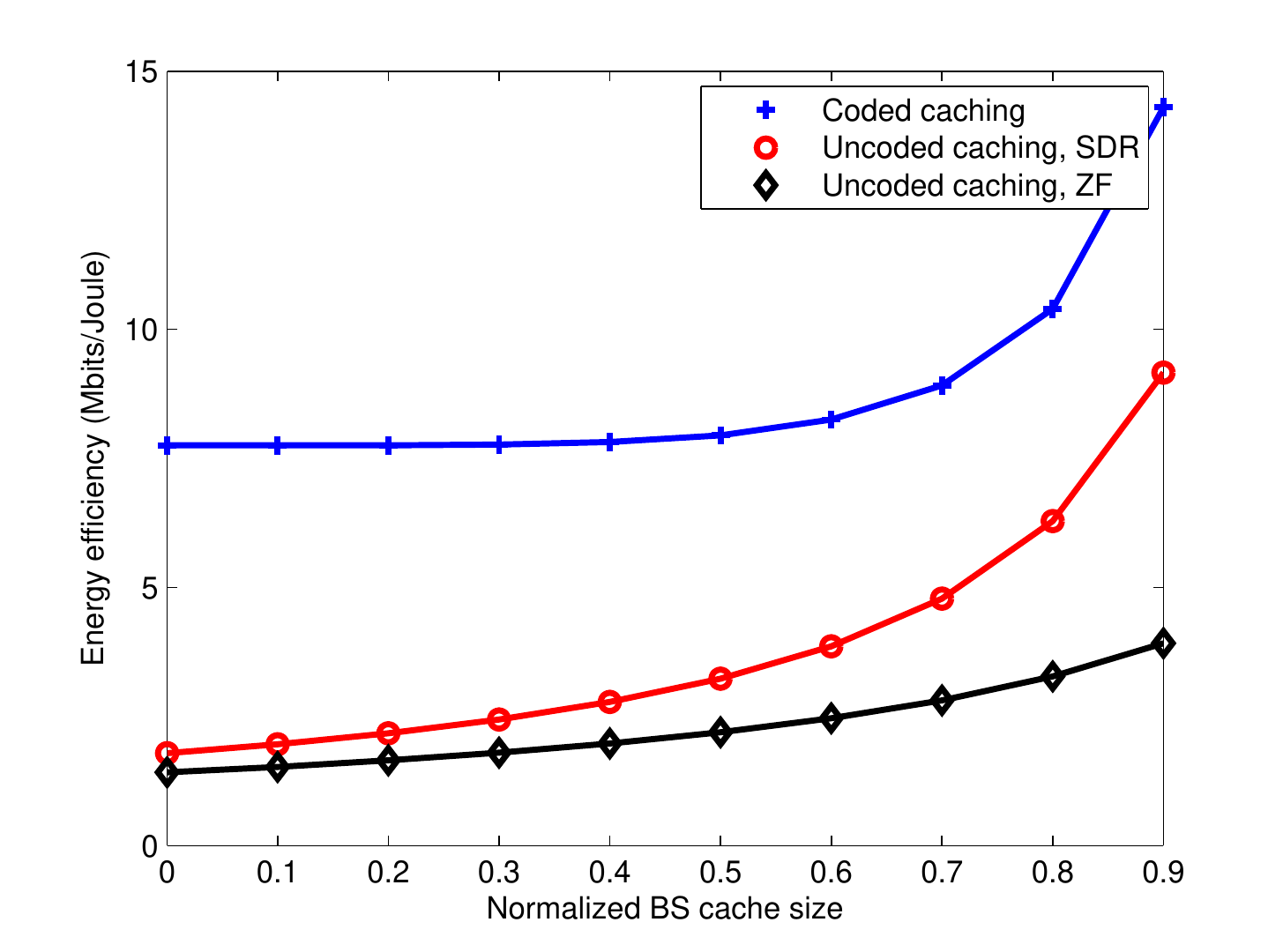}}
	\end{center}
	\caption{Energy efficiency of the two caching methods. (a) - EE v.s. normalized user cache size, cost-free on backhaul; (b) - EE v.s. normalized user cache size; (c) - EE v.s. normalized user cache size with small values (d) - EE v.s. normalized BS cache size.}\label{fig:eta}
\end{figure*}
\section{Numerical results}\label{sec:Results}
This section presents numerical results to demonstrate the effectiveness of the studied caching policies. The results are averaged over 500 channel realizations. For ease of presentation, the uncoded caching under the general beamformer design using SDR in Section~\ref{sec:EE unc SDR} is named as \emph{SDR} and the Zero-forcing design in Section~\ref{sec:EE unc ZF} is named as \emph{ZF} in the figures. Unless otherwise stated, the system setup is as follows: $L = 10$ antennas, $K = 8$ users, $N = 1000$ files, $B = 1$ MHz, $\eta = 10^{-6}$ bits/Joule \cite{TaoCheZhoYu2016}, $\sigma^2_{h_k} = 1, \forall k$, $Q = 10$ Mb, $\gamma_k = 2\ \text{Mbps}~, \forall k$.

\subsection{Energy efficiency performance}
We first study the two caching strategies when the energy consumption on the backhaul is negligible. This occurs when the BS cache is large enough to store all the files. In this case, the EE only depends on the user cache size. Figure~\ref{fig:eta}a presents the EE of the two caching strategies as the function of the normalized user cache size (the user cache size $M_u$ divided by the library size $N$). The EE is plotted based on the optimal precoding vectors obtained from Section~\ref{sec:EE maximization}. It is shown that the uncoded caching under the SDR design achieves higher EE than the coded caching when the normalized user cache is less than 0.2. This result suggests an important guideline for using the uncoded caching since the user cache is usually small compared to the library size in practice. When the user cache is capable of storing more than 20\% of all the files, it suggests to use the coded caching for larger system EE. It is also observed that the uncoded caching under SDR design achieves higher EE than the ZF for all user cache size. This is because the SDR design is more efficient than the ZF precoding. 

Figure~\ref{fig:eta}b compares the EE for various user cache size when $M_b = 0.7N$. In general, the coded caching method is more efficient than the uncoded caching for most of user cache size values. Increasing user cache capability results in larger relative gain of the coded-caching compared with the uncoded method. The uncoded caching under SDR design achieves slightly better EE than the ZF design at small user cache sizes, however, at an expense of higher computational complexity as shown in Table~\ref{table:3}. From the practical point of view, ZF design is preferred in this case because of its low complexity. When $M_b$ increases, the SDR achieves significantly higher EE than the ZF. Figure~\ref{fig:eta}c presents the EE v.s. the user cache size when both BS and user cache size are small. It is shown that the uncoded caching strategy with either SDR or ZF design outperforms the coded caching scheme in the observed user cache sizes, which is in line with the result in Figure~\ref{fig:eta}b. Figure~\ref{fig:eta}d compares the EE as a function of the BS cache size when $M_u = 0.5N$. The result shows that the caching at the BS has more impacts on both the caching strategies when the BS cache size is relatively large. It is shown that the coded-caching outperforms the uncoded caching for all values $M_b$. It is also shown that the SDR design achieves higher EE gain compared with the ZF as $M_b$ increases.

\begin{figure}
	\centering 
	\subfigure[$M_b = 0.3N$]{\includegraphics[width=\columnwidth]{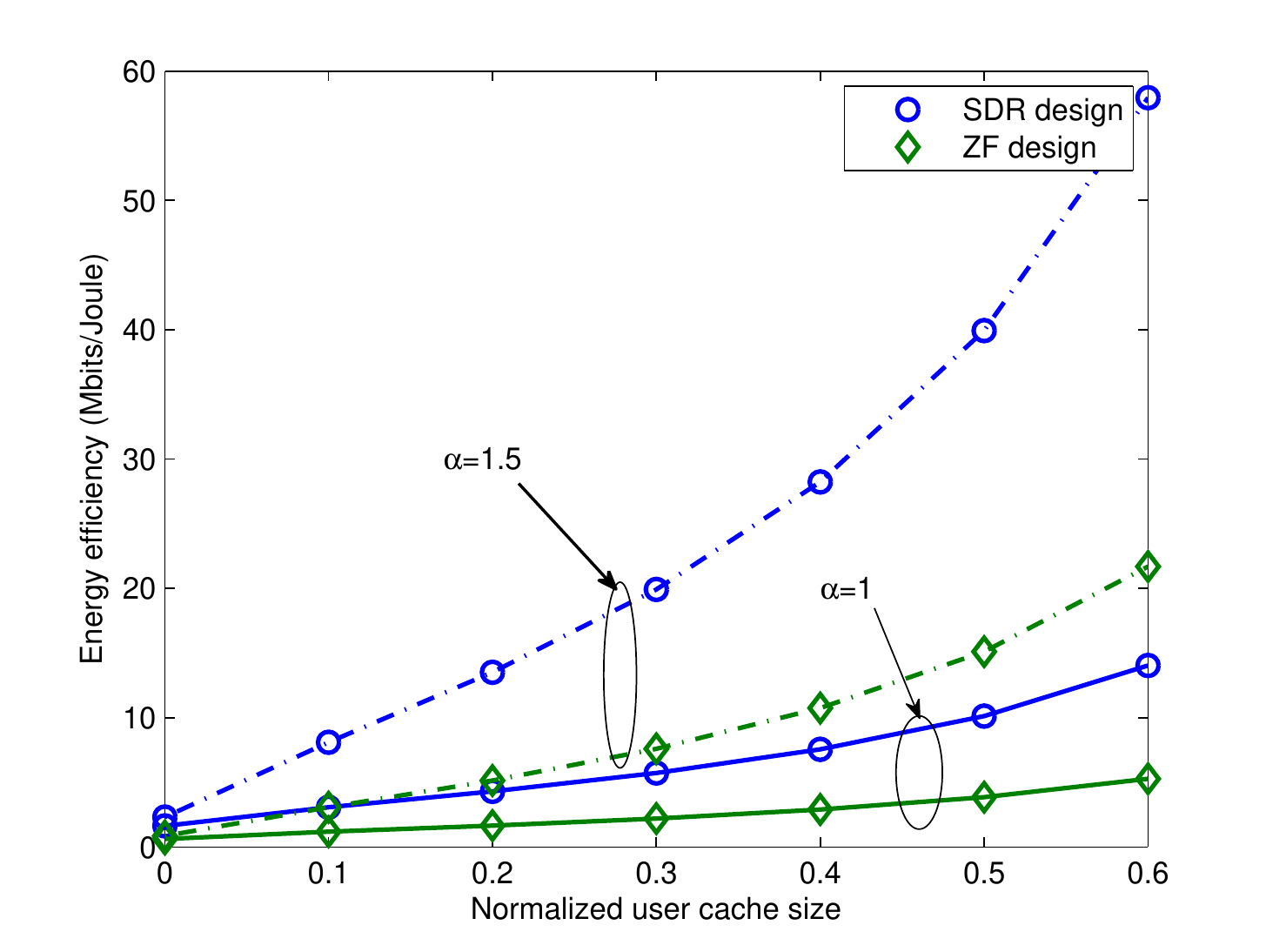}}\vspace{-0.3cm}
	\subfigure[$M_u = 0.3N$]{\includegraphics[width=\columnwidth]{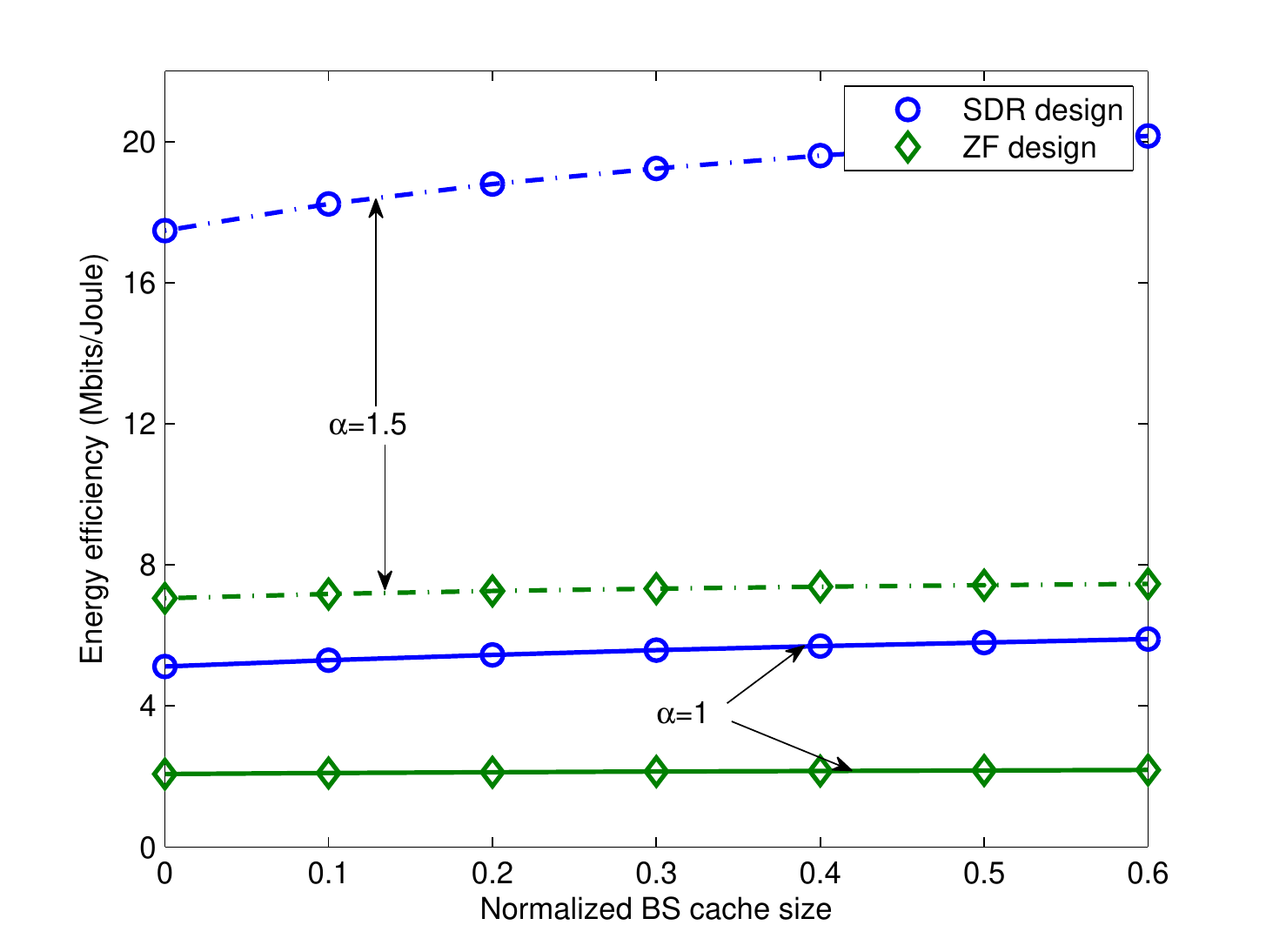}}
	\caption{Energy efficiency of the uncoded caching algorithm with Zipf content popularity distribution with different Zipf exponents.}\label{fig:Zipf}
\end{figure}

Figure~\ref{fig:Zipf}a presents the EE v.s. the normalized user cache size of the uncoded caching algorithm under Zipf content popularity distribution, i.e., $q_{k,n} = \frac{n^{-\alpha}}{\sum_{i=1}^N i^{-\alpha}}, \forall k$. It is observed that the SDR design significantly surpasses the ZF design. In particular, at 40\% library size of the user cache, the SDR achieves almost 3 times EE higher than the ZF design. Greater Zipf exponent factor results in higher EE for the both designs. This is because the content distribution in this case is more centralized at some files. Figure~\ref{fig:Zipf}b plots the EE v.s. the normalized BS cache size. Similarly, the SDR design achieves higher EE than the ZF design. Also, the BS cache size has smaller impacts on the system EE than the user cache size.

\subsection{Delivery time performance}
\begin{figure}
	\centering 
	\includegraphics[width=\columnwidth]{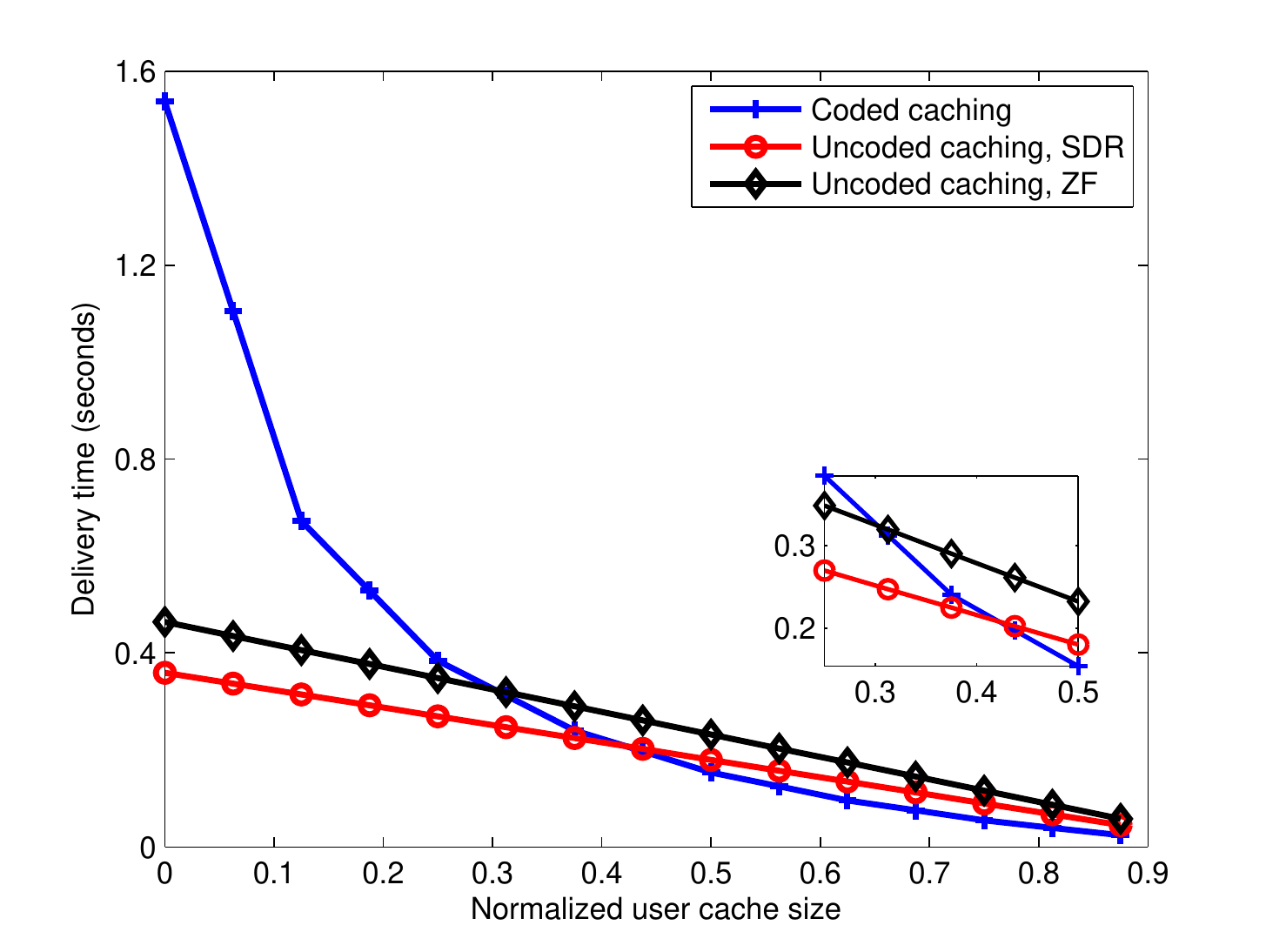}
	\caption{Delivering time of the two caching methods v.s. the normalized user memory $M_u$. Average transmit power is 10 dB.}\label{fig:TimevsLBS}
	\vspace{-0.3cm}
\end{figure}

Figure~\ref{fig:TimevsLBS} presents the delivery times of the two caching strategies as a function of the user cache size with 8 users and transmit power equal to 10 dB. It is shown that the uncoded caching strategy with both designs outperforms the coded counter part if the user cache is smaller than 30\% of the library. When the cache size is larger, the coded-caching method achieves slightly smaller latency than the uncoded caching strategy. This important observation suggests the optimal caching algorithm in practical systems depending on the memory availability at the edge nodes. It is also shown that the delivery time of the uncoded caching strategy linearly depends on the cache size. This can be seen from Proposition~\ref{prop:1} that the network throughput in the uncoded caching linearly depends on the cache size. 
\begin{figure}
	\centering 
	\subfigure[$M_u = 0.3N$]{\includegraphics[width=\columnwidth]{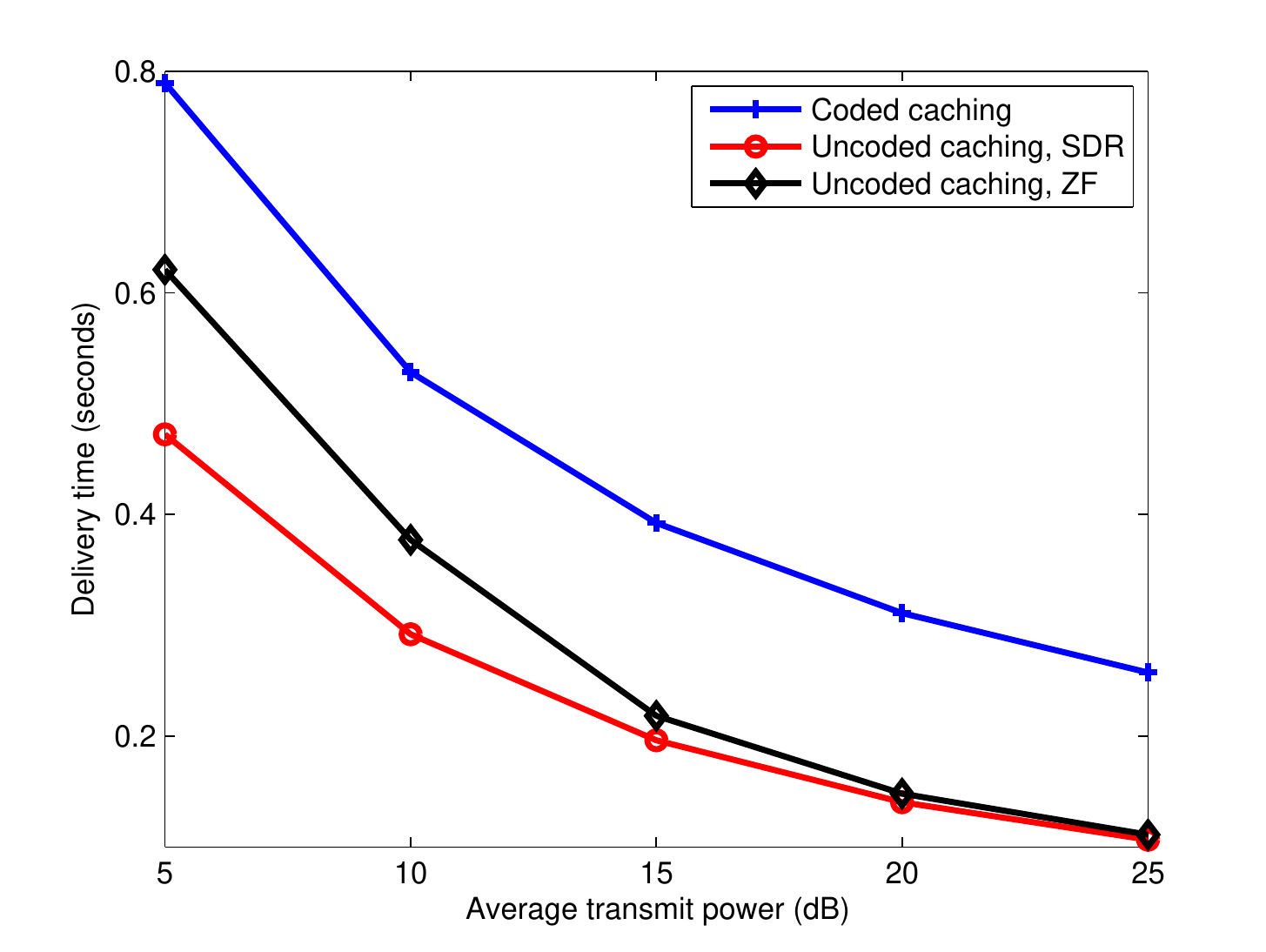}}\vspace{-0.35cm}
	\subfigure[$M_u = 0.7N$]{\includegraphics[width=\columnwidth]{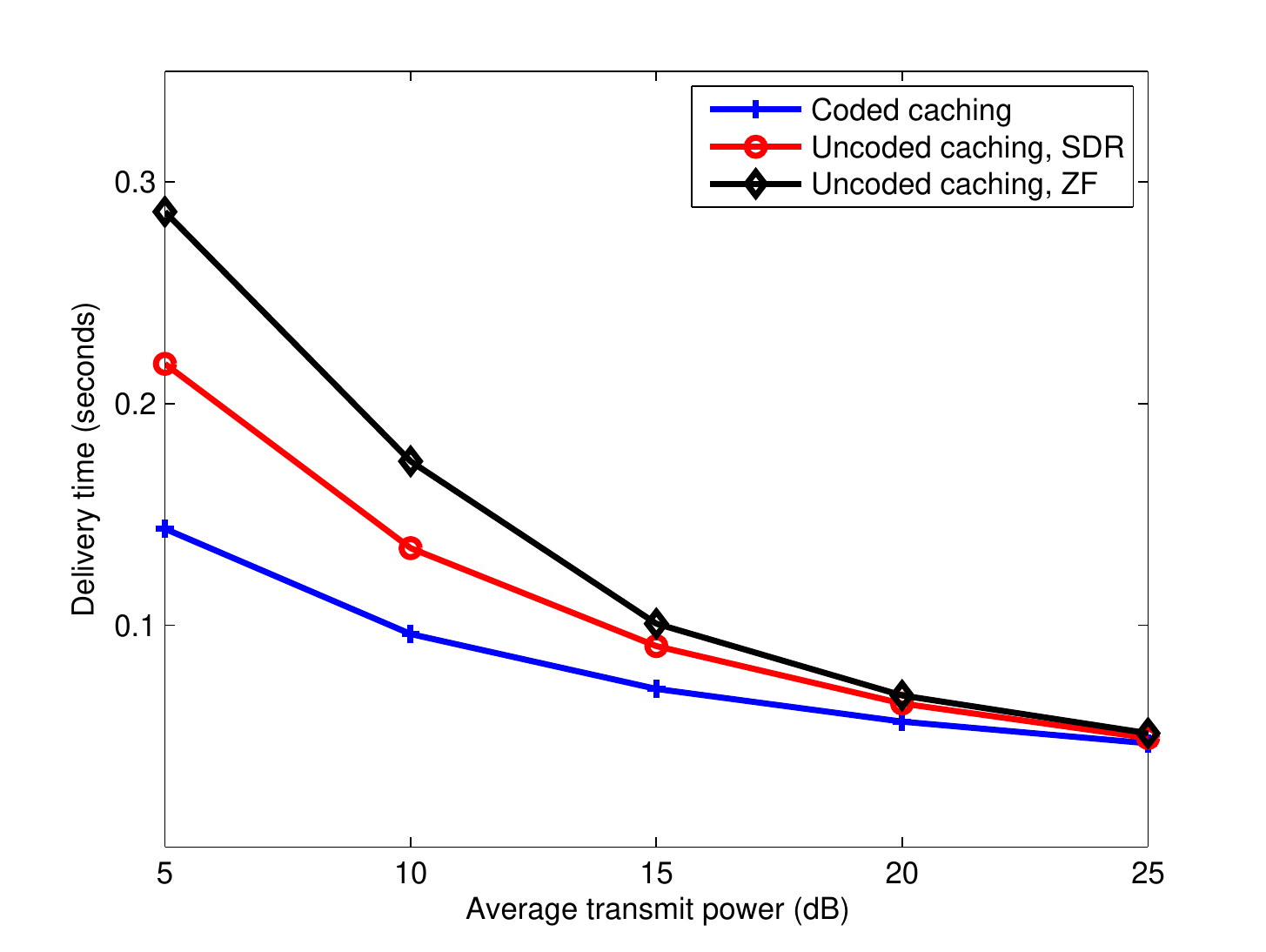}}	
	\caption{Delivering time of the two caching methods v.s. the average transmit power. }\label{fig:TimevsPsum}
\end{figure}
\begin{figure}
	\centering 
	\includegraphics[width=\columnwidth]{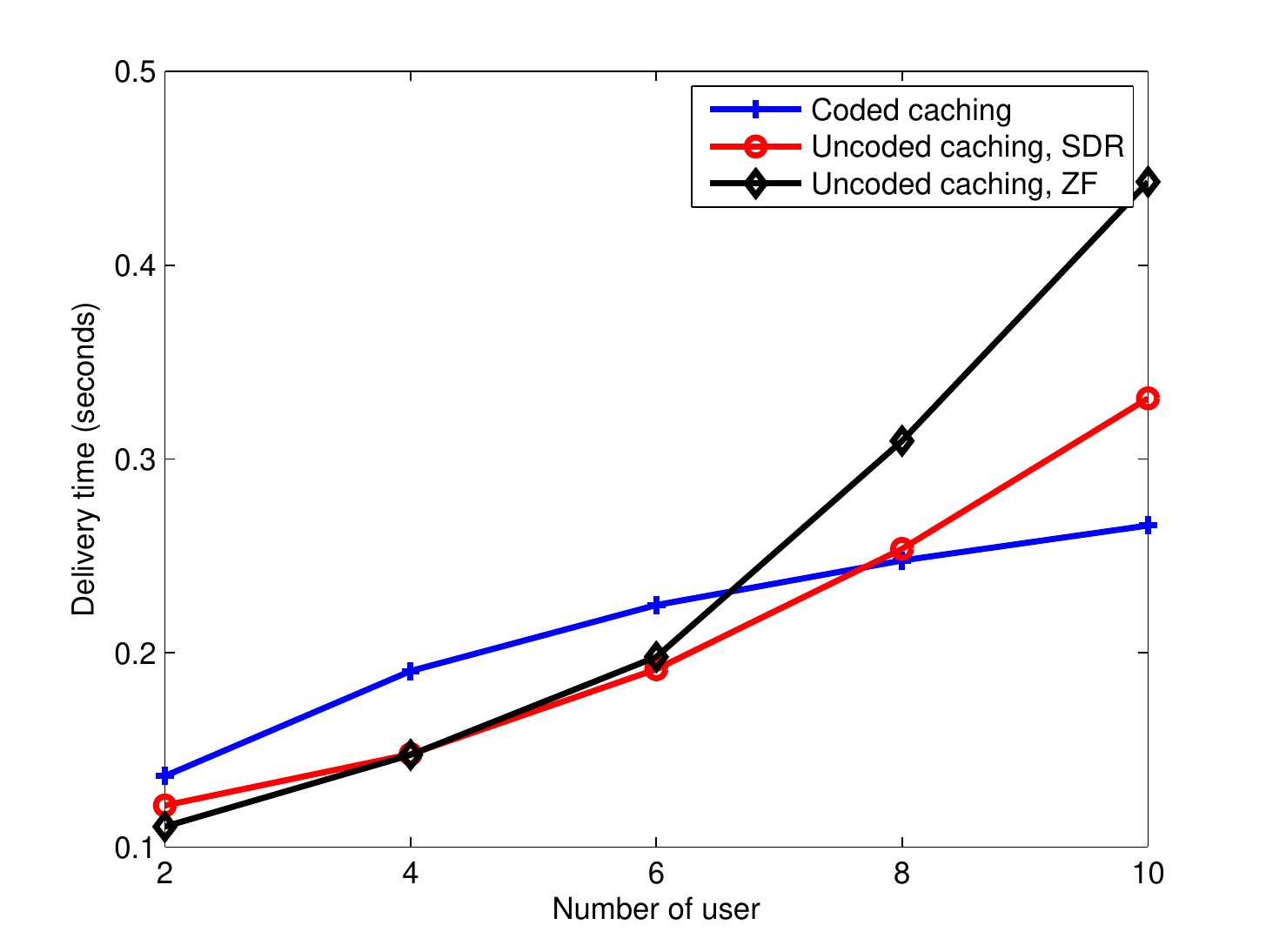}
	\caption{Delivering time of the two caching methods v.s. the number of users. Average transmit power is 10 dB, $M_u = 0.4N$.}\label{fig:TimevsK}
\end{figure}

Figure~\ref{fig:TimevsPsum} compares the delivery times of the two caching algorithms for various transmit powers. Obviously, increasing the transmit power will significantly reduce the delivery times in both strategies. When the user cache size is small (Fig.~\ref{fig:TimevsPsum}a), the uncoded caching strategies deliveries the requested files faster than the coded caching method, which is in line with the results in Fig.~\ref{fig:TimevsLBS}. When the user cache memory is capable of storing more content (Fig.~\ref{fig:TimevsPsum}b), the coded caching strategy is more efficient than the uncoded caching. It is also observed that the SDR design only outperforms the ZF design for small transmit power. This is because large transmit power can supports optimal solution for both SDR and ZF designs. 

Figure~\ref{fig:TimevsK} plots the delivery times depending on the number of users $K$. For small $K$, the uncoded caching strategy slightly outperforms the coded caching method. When $K$ increases, the coded caching tends to surpass the uncoded caching strategy. In this case, the total cache size in the network is bigger in which the coded caching algorithm is more effective. 

	\section{Conclusions} \label{sec:Conclusions}
	We have analysed the performance of cache-assisted wireless networks under two notable uncoded and coded caching strategies. First, we have expressed the energy efficiency metric in closed-form expression for each caching strategy as a function of base station and user cache sizes and the transmit power on the access links. Based on the derived closed-form, two optimization problems have been formulated to maximize the system EE while satisfying a predefined user rate requirement. Second, we have analysed the total delivery time for each caching strategy and designed the beamforming vectors to minimize the total delivery time. It has been shown that the uncoded caching algorithm achieves higher EE than the coded caching method only when the user cache size is small and the BS cache is large enough.	
	
	Based on the studied work, several research directions can be extended. One is to consider generic networks in which the data centre is serving multiple base stations. In this case, different backhaul constraints for each BS should be taken into account when designing the caching algorithms. Another direction is to consider the coded caching algorithm applied to non-uniform content popularity. This requires a redesign of both cache placement and delivery phases in order to take into consideration differences in user preferences.
\balance
\appendices

\section{Proof of Proposition~\ref{prop:1}}\label{app:1}
The proof can be found by similar techniques in \cite[Sec. II]{Vu2017}. When a user requests a file, parts of the requested file are in the user cache. Since the users' requests are independent, the requested files can be either the same or different.

For any integer number $m, 1 \leq m \leq N$, there are $N^{m}$ ways to choose $m$ elements out of the set of size $N$, which can be further expressed as 
\begin{align*}
N^{m} = \sum_{l=1}^{m} a^{m}_l \mathcal{C}^N_l,
\end{align*}
where $\mathcal{C}^N_l \triangleq \frac{N!}{(N-l)!}$ and $a^{m}_l$ is a constant. In the above equation, $a^{m}_l \mathcal{C}^N_l$ is the number of choices of $m$ elements out of $N$ which contains $l$ different elements. By using the inductive method, we can obtain:
\begin{align*}
a^m_l = \left\{ \begin{array}{ll}
1,&\ \text{if}\ l = 1\ \text{or}\ m \\
ma^{m-1}_l + a^{m-1}_{l-1},&\ \text{if}\ 1 < l < m	
\end{array}\right.
\end{align*}
For a choice comprising of $l$ different values, the BS needs to send $lQ(1 - M_u/N)$ subfiles to the users. Therefore, the average access throughput is calculated as
\begin{align}
Q_{\rm unc,AC} &= \frac{1}{N^{K}} \sum_{l=1}^{K} l Qa^{K}_l \mathcal{C}^N_l \left(1 - \frac{M_u}{N}\right) \notag\\
&= \sum_{l=1}^{K} \frac{l Q a^{K}_l}{N^{K-l}} \left(1 - \frac{M_u}{N}\right) \prod_{i=1}^l \frac{N-l+i}{N}. \label{eq: prop 11}		
\end{align}
It is observed that the library size $N$ is usually very large compared to $K$, thus $\frac{N-l+i}{N} \simeq 1, \forall 1\leq i \leq l$ and 
\begin{align}
\frac{la^{K}_l}{N^{K - l}} \simeq \left\{ 
\begin{array}{ll}
0,~\text{if}~l < K\\
K,~\text{if}~l = K
\end{array}.
\right. \label{eq: prop 22}
\end{align}
From \eqref{eq: prop 11} and \eqref{eq: prop 22} we obtain:
\begin{align}
Q_{\rm unc,AC} \simeq KQ\left( 1 - \frac{M_u}{N}\right). 
\end{align}  

To compute the backhaul throughput, we note that the BS randomly cache $\frac{M_b}{N}$ parts of every file. Therefore, the probability that a bit is stored at the BS cache is $\frac{M_b}{N}$. Finally, since the BS is the caching at the BS and users independent, we obtain $Q_{\rm unc, BH}$ in Proposition 1.


\begin{thebibliography}{10}
	\providecommand{\url}[1]{#1}
	\csname url@samestyle\endcsname
	\providecommand{\newblock}{\relax}
	\providecommand{\bibinfo}[2]{#2}
	\providecommand{\BIBentrySTDinterwordspacing}{\spaceskip=0pt\relax}
	\providecommand{\BIBentryALTinterwordstretchfactor}{4}
	\providecommand{\BIBentryALTinterwordspacing}{\spaceskip=\fontdimen2\font plus
		\BIBentryALTinterwordstretchfactor\fontdimen3\font minus
		\fontdimen4\font\relax}
	\providecommand{\BIBforeignlanguage}[2]{{%
			\expandafter\ifx\csname l@#1\endcsname\relax
			\typeout{** WARNING: IEEEtran.bst: No hyphenation pattern has been}%
			\typeout{** loaded for the language `#1'. Using the pattern for}%
			\typeout{** the default language instead.}%
			\else
			\language=\csname l@#1\endcsname
			\fi
			#2}}
	\providecommand{\BIBdecl}{\relax}
	\BIBdecl
	

	\bibitem{Cisco}
	Cisco, ``Cisco visual networking index: Global mobile data traffic forecast
	update 2016-2021,'' 2017, white paper.
	
	\bibitem{Vu2016}
	T.~X. Vu, H.~D. Nguyen, T.~Q.~S. Quek, and S.~Sun, ``Adaptive cloud radio
	access networks: compression and optimization,'' \emph{IEEE Trans. Signal
		Process}, vol.~65, no.~1, pp. 228--241, Jan. 2017.
	
	\bibitem{Tran1}
	T. X. Tran and D. Pompili, ``Dynamic Radio Cooperation for User-Centric Cloud-RAN With Computing Resource Sharing,'' \emph{IEEE Trans. Wireless Commun.}, vol.~16, no.~4, pp. ~2379--2393, Apr. 2017.
	
	\bibitem{Tran2}
	T. X. Tran, A. Hajisami, and D. Pompili, ``QuaRo: A queue-aware robust coordinated transmission strategy for downlink C-RANs,'' in \emph{Proc. IEEE Int. Conf. Sensing, Commun. and Netw.}, London, 2016, pp. 1-9.
	
	\bibitem{Borst2010}
	S.~Borst, V.~Gupta, and A.~Walid, ``Distributed caching algorithms for content
	distribution networks,'' in \emph{Proc. IEEE Int. Conf. Comput. Commun.},
	Mar. 2010, pp. 1--9.
	
	\bibitem{AliNie2014}
	M.~A. Maddah-Ali and U.~Niesen, ``Fundamental limits of caching,'' \emph{IEEE
		Trans. Inf. Theory}, vol.~60, no.~5, pp. 2856--2867, May 2014.
	
	\bibitem{Vu2017}
	T.~X. Vu, S.~Chatzinotas, and B.~Ottersten, ``Coded caching and storage allocation in heterogeneous networks,'' in \emph{Proc. IEEE Wireless Commun.
			Netw. Conf.}, San Francisco, CA, 2017, pp. 1--5.
		
	\bibitem{Almeroth1996}
	K.~C. Almeroth and M.~H. Ammar, ``The use of multicast delivery to provide a
	scalable and interactive video-on-demand service,'' \emph{IEEE J. Sel. Areas
		Commun.}, vol.~14, no.~6, pp. 1110--1122.
	
	\bibitem{Christop2015d}
	D.~Christopoulos, S.~Chatzinotas, and B.~Ottersten, ``Cellular-broadcast
	service convergence through caching for \uppercase{comp} cloud
	\uppercase{RAN},'' in \emph{Proc. IEEE Symp. Commun. Veh. Tech. in the
		Benelux}, Luxembourg, 2015, pp. 1--6.
	
	\bibitem{JiCaire2016}
	M.~Ji, G.~Caire, and A.~F. Molisch, ``Fundamental limits of caching in wireless
	\uppercase{D2D} networks,'' \emph{IEEE Trans. Inf. Theory}, vol.~62, no.~2,
	pp. 849--869, Feb. 2016.
	
	\bibitem{Sengupta2016}
	A.~Sengupta, R.~Tandon, and T.~C. Clancy, ``Fundamental limits of caching with
	secure delivery,'' \emph{IEEE Trans. Info. Forensics and Security}, vol.~10,
	no.~2, pp. 355--370, Feb. 2015.
	
	\bibitem{Sengupta2016b}
	A.~Sengupta, R.~Tandon, and O.~Simeone, ``Cache aided wireless networks: Tradeoffs between storage and latency,'' in \emph{Proc. Annu. Conf. Info.
		Sci. Syst.}, Princeton, NJ, Mar. 2016, pp. 320--325.
	
	\bibitem{SHPark2017}
	S.~H.~Park, O.~Simeone, W.~Lee, and S.~Shamai, ``Coded multicast fronthauling and edge caching for multi-connectivity transmission in fog radio access networks,'' in \emph{Proc. IEEE Int. Workshop Signal Process. Adv. Wireless Commun.}, Sapporo, Japan, 2017, pp. 1-5.
	
	\bibitem{Karamchandani2016}
	N.~Karamchandani, U.~Niesen, M.~A. Maddah-Ali, and S.~N. Diggavi,
	``Hierarchical coded caching,'' \emph{IEEE Trans. Inf. Theory}, vol.~62,
	no.~6, pp. 3212--3229, Jun. 2016.
	
	\bibitem{Tang2016}
	L.~Tang and A.~Ramamoorthy, ``Coded caching for networks with the resolvability
	property,'' in \emph{Proc. IEEE Int. Symp. Inf. Theory}, Barcelona, Jul.
	2016, pp. 420--424.
	
	\bibitem{TaoCheZhoYu2016}
	M.~Tao, E.~Chen, H.~Zhou, and W.~Yu, ``Content-centric sparse multicast
	beamforming for cache-enabled cloud \uppercase{RAN},'' \emph{IEEE Trans.
		Wireless Commun.}, vol.~15, no.~9, pp. 6118--6131, Sept. 2016.
	
	\bibitem{Vu2018a}
		T.~X. Vu, S. Chatzinotas, and B. Ottersten ``Energy Minimization for Cache-assisted Content Delivery Networks with Wireless Backhaul,'' \emph{IEEE Wireless Commun. Lett.}, vol.~pp, no.~pp, pp. 1--1, 2018.
		
	\bibitem{Khreishah2016}
	A.~Khreishah, J.~Chakareski, and A.~Gharaibeh, ``Joint caching, routing, and
	channel assignment for collaborative small-cell cellular networks,''
	\emph{IEEE J. Sel. Areas Commun.}, vol.~34, no.~8, pp. 2275--2284, IEEE
	Trans. Inf. Theory. 2016.
	
	\bibitem{ZhaXiaWuLi2016}
	L.~Zhang, M.~Xiao, G.~Wu, and S.~Li, ``Efficient scheduling and power
	allocation for \uppercase{D2D}-assisted wireless caching networks,''
	\emph{IEEE Trans. Commun.}, vol.~64, no.~6, pp. 2438--2452, Jun. 2016.
	
	\bibitem{Gregori2016}
	M.~Gregori, J.~Gómez-Vilardebó, J.~Matamoros, and D.~Gündüz, ``Wireless
	content caching for small cell and \uppercase{D2D} networks,'' \emph{IEEE J.
		Sel. Areas Commun.}, vol.~34, no.~5, pp. 1222--1234, May 2016.
	
	\bibitem{Ji2016}
	M.~Ji, G.~Caire, and A.~F. Molisch, ``Wireless device-to-device caching
	networks: Basic principles and system performance,'' \emph{IEEE J. Sel. Areas
		Commun.}, vol.~34, no.~1, pp. 176--189, Jan. 2016.
	
	\bibitem{Ji2015a}
	M.~Ji, G.~Caire, and A.~Molisch, ``The throughput-outage tradeoff of wireless
	one-hop caching networks,'' \emph{IEEE Trans. Inf. Theory}, vol.~61, no.~12,
	pp. 6833--6859, Dec. 2015.
	
	\bibitem{Yang2016}
	C.~Yang, Y.~Yao, Z.~Chen, and B.~Xia, ``Analysis on cache-enabled wireless
	heterogeneous networks,'' \emph{IEEE Trans. Wireless Commun.}, vol.~15,
	no.~1, pp. 131--145, Jan. 2016.
	
	\bibitem{Chen2016}
	Z.~Chen, J.~Lee, T.~Q. Quek, and M.~Kountouris, ``Cooperative caching and
	transmission design in cluster-centric small cell networks,'' \emph{IEEE
		Trans. Wireless Commun.}, vol.~16, no.~5, pp. 3401 -- 3415, May 2016.
	
	\bibitem{Alfano2016}
	G.~Alfano, M.~Garetto, and E.~Leonardi, ``Content-centric wireless networks
	with limited buffers: when mobility hurts,'' \emph{IEEE/ACM Trans. Netw.},
	vol.~24, no.~1, pp. 299--311, Jan. 2016.
	
	\bibitem{Tran2016}
	T. X. Tran, F. Kazemi, E. Karimi, and D. Pompili, ``Mobee: Mobility-aware energy-efficient coded Caching in cloud radio access networks,'' in \emph{Proc. IEEE Int. Conf. Mobile Ad-Hoc Sensor Syst. (MASS)}, Orlando, FL, 2017, pp. 461--465.
	
	\bibitem{Gabry2016}
	F.~Gabry, V.~Bioglio, and I.Land, ``On energy-efficient edge caching in
	heterogeneous networks,'' \emph{IEEE J. Sel. Areas Commun.}, vol.~34, no.~12,
	pp. 3288--3298, Dec. 2016.
	
	\bibitem{Liu2016}
	D.~Liu and C.~Yang, ``Energy efficiency of downlink networks with caching at
	base stations,'' \emph{IEEE J. Sel. Areas Commun.}, vol.~34, no.~4, pp.
	907--922, Apr. 2016.
	

		
	\bibitem{Vu2017a}
	T.~X. Vu, S.~Chatzinotas, and B.~Ottersten, ``Energy-efficient design for
	edge-caching wireless networks: When is coded-caching beneficial?'' in
	\emph{Proc. IEEE Int. Workshop Signal Process. Wireless Commun.}, Sapporo,
	2017, pp. 1--5.
	
	\bibitem{Sidirop2006}
	N.~D. Sidiropoulos, T.~N. Davidson, and Z.-Q. Luo, ``Transmit beamforming for
	physical-layer multicasting,'' \emph{IEEE Trans. Signal Process}, vol.~54,
	no.~6, pp. 2239--2251, Jun. 2006.
	
	\bibitem{Luo2010}
	Z.-Q. Luo, W.~K. Ma, A.~M.~C. So, Y.~Ye, and S.~Zhang, ``Semidefinite
	relaxation of quadratic optimization problems,'' \emph{IEEE Signal Process.
		Mag.}, vol.~27, no.~3, pp. 20--34, Mar. 2010.
	
	\bibitem{Boyd2004}
	S.~Boyd and L.~Vandenberghe, \emph{Convex Optimization}.\hskip 1em plus 0.5em
	minus 0.4em\relax Cambridge Univ. Press, 2004.
	
\end{thebibliography}
\end{document}